\pdfoutput=1
\documentclass[aps,prd,onecolumn,superscriptaddress,preprintnumbers,nofootinbib, a4paper]{revtex4}

\usepackage[utf8x]{inputenc}
\usepackage{amssymb,amsmath,mathrsfs,enumerate}
\usepackage{graphicx,rotate, multirow}
\usepackage{braket}
\usepackage[normalem]{ulem}
\usepackage{wrapfig}
\usepackage[textfont=it,labelfont=bf]{caption}
\usepackage{bm}
\usepackage[colorlinks=true,
linkcolor=red,
urlcolor=blue,
citecolor=blue]{hyperref}
\usepackage{lineno}
\usepackage{color}
\usepackage{bbold}
\usepackage{xcolor}
\usepackage{fancyhdr}
\usepackage{geometry}
\usepackage{float}
\long\def\rpl#1!!#2!!{\textcolor{red}{#1} \textcolor{blue}{#2}}

\usepackage{subcaption}

\makeatother

\newcommand{\bea}{\begin{eqnarray}}
\newcommand{\eea}{\end{eqnarray}}

\newcommand{\be}{\begin{equation}}
\newcommand{\ee}{\end{equation}}
\def\beq{\begin{equation}}
\def\eeq{\end{equation}}
\newcommand{\ba}{\begin{eqnarray}}
\newcommand{\ea}{\end{eqnarray}}
\def\ifmath#1{\relax\ifmmode #1\else $#1$\fi}
\def\eq#1{Eq.~(\ref{#1})}
\def\eqs#1#2{Eqs.~(\ref{#1}) and (\ref{#2})}

\renewcommand{\Re}{\textrm{Re}}
\renewcommand{\Im}{\textrm{Im}}

\usepackage{physics}
\newcommand{\one}{\mathbb{1}}
\allowdisplaybreaks

\newcommand{\AddrCFTP}{%
Departamento de Física and CFTP, Instituto Superior Técnico\\
Universidade de Lisboa, 
Av.\ Rovisco Pais 1, 1049-001 Lisboa, Portugal }

\begin{document}

\title{Surveying the complex three Higgs doublet model with Machine Learning} 
\author{Rafael Boto}\email{rafael.boto@tecnico.ulisboa.pt}
\affiliation{\AddrCFTP}
\author{João A. C. Matos}\email{joao.a.costa.matos@tecnico.ulisboa.pt}
\affiliation{\AddrCFTP}
\author{Jorge C.~Romão}\email{jorge.romao@tecnico.ulisboa.pt}
\affiliation{\AddrCFTP} 
 \author{João  P.~Silva} \email{jpsilva@cftp.ist.utl.pt}
 \affiliation{\AddrCFTP}
\renewcommand*{\thefootnote}{\arabic{footnote}}
\setcounter{footnote}{0}

\begin{abstract}
The couplings of the $125\,\textrm{GeV}$ Higgs are being measured with higher
precision as the Run 3 stage of LHC continues.
Models with multiple Higgs doublets allow potential deviations from the SM predictions.
For more than two doublets,
there are five possible types of models that avoid flavor changing neutral
couplings at tree level by the addition of a symmetry.
We consider a softly broken $\mathbb{Z}_2\times\mathbb{Z}_2$
three-Higgs doublet model with explicit CP violation in the scalar
sector,
exploring all five possible types of coupling choices and all five mass
orderings of the neutral scalar bosons.
The phenomenological study is performed using a Machine Learning
black box optimization algorithm that efficiently searches for
the possibility of large pseudoscalar Yukawa couplings.
We identify the model choices that allow a purely pseudoscalar
coupling in light of all recent experimental limits,
including direct searches for CP-violation,
thus motivating increased effort into improving the
experimental precision.
\end{abstract}

\maketitle

\section{\label{sec:intro}Introduction}

The scalar sector lies at the heart of many outstanding
problems in particle physics:
what is the source of the mass and mixing features in flavour?;
does the solution to the dark matter (DM) problem involve scalar DM or
a scalar portal?;
what are the phase transitions and/or new sources of CP violation
required for baryogenesis?
The 2012 discovery of the $125\textrm{GeV}$ Higgs boson ($h_{125}$) by
ATLAS \cite{ATLAS:2012yve} and CMS \cite{CMS:2012qbp}
was only the start of this major effort.
Generically, one is now searching for new particles and also probing ever more incisively
all couplings and features of $h_{125}$.

In particular, the CP nature of the ($h_{125}$) couplings to fermions
only recently became under experimental scrutiny,
with bounds on the CP-odd ($c^o_{tt}$) versus CP-even ($c^e_{tt}$) 
couplings to the top quark \cite{ATLAS:2020ior}:
\be
|\theta_{t}| = |\arctan(c^o_{tt} / c^e_{tt})| < 43^\circ
\ \ \textrm{at}\ \ 
95\%
\ \ \textrm{CL}.
\label{theta_t}
\ee
This bound was obtained by looking for $t \bar{t} h_{125}$ and $t h_{125}$ processes,
using $h_{125} \rightarrow \gamma\gamma$ with the ATLAS detector.
More recently,
bounds were placed on the CP-odd ($c^o_{\tau\tau}$) versus CP-even ($c^e_{\tau\tau}$) 
couplings to the tau lepton \cite{CMS:2021sdq,ATLAS:2022akr}:
\be
|\theta_{\tau}| = |\arctan(c^o_{\tau\tau} / c^e_{\tau\tau})| < 34^\circ
\ \ \textrm{at}\ \ 
95\%
\ \ \textrm{CL}.
\label{theta_tau}
\ee
Both experiments \cite{CMS:2021sdq,ATLAS:2022akr} use the angular correlation
between the decay (leptonic and/or hadronic) planes of $\tau$ leptons produced
in Higgs boson decays.
There are currently no direct bounds on the CP-odd coupling to the bottom quark
($c^o_{bb}$).

Models with extra scalars and, in particular, N Higgs doublet models (NHDM)
have long be studied as possible solutions to the Standard Model (SM) shortcomings.
For example,
adding one extra scalar doublet to the SM permits the appearance of
new sources of CP violation, allowing for spontaneous CP violation \cite{Lee:1973iz}
or for successful baryogenesis \cite{McLerran:1988bb,Turok:1990in,Turok:1990zg,Cohen:1991iu}.
Such 2HDM have been analyzed extensively; for reviews, see, for example
Refs.~\cite{Gunion:1989we,Branco:2011iw}.
A simple scenario with CP violation in 2HDM consists in using a complex parameter
to break softly a $\mathbb{Z}_2$ symmetry; a model known as 
the complex two Higgs doublet model (C2HDM),
first introduced in \cite{Ginzburg:2002wt}
and much explored subsequently,
for example in \cite{Khater:2003wq,
ElKaffas:2007rq, Grzadkowski:2009iz, Arhrib:2010ju, Barroso:2012wz,Abe:2013qla,
Inoue:2014nva, Cheung:2014oaa, Fontes:2014xva,Grober:2017gut,Fontes:2015xva, Fontes:2015mea,
Chen:2015gaa, Chen:2017com, Muhlleitner:2017dkd, Fontes:2017zfn, Basler:2017uxn,
Aoki:2018zgq, Basler:2019iuu, Wang:2019pet, Boto:2020wyf, Cheung:2020ugr,
Altmannshofer:2020shb, Fontes:2019wqh, Fontes:2021iue, Basler:2021kgq, Frank:2021pkc, Abouabid:2021yvw,
Fontes:2022izp,Azevedo:2023zkg,Goncalves:2023svb,Biekotter:2024ykp}. 

Within the framework of the C2HDM,
a notably unusual scenario was identified in \cite{Fontes:2015mea}:
$h_{125}$ could exhibit a purely scalar coupling to top quarks
and a purely pseudoscalar coupling to bottom quarks.
This was still consistent with the experimental data available
at the end of 2017 \cite{Fontes:2017zfn}.
In a stark illustration of the strong experimental effort
since then, 
by 2024,
more data on $h_{125}$,
direct searches for CP-violation in $c^o_{tt}$ and $c^o_{\tau\tau}$,
new searches for heavier scalars,
and, especially,
new results on the electron electric dipole moment,
excluded this scenario \cite{Biekotter:2024ykp}.
This possibility was partly resuscitated by Ref.~\cite{Boto:2024jgj},
in a three Higgs doublet with explicit CP violation (C3HDM),
with very specific characteristics:
each doublet couples to a different charged fermion sector (a so-called Type Z model),
and $h_{125}$ is identified as the lightest scalar particle.

The results mentioned thus far were obtained with what one could call
informed traditional simulation techniques.
Indeed, blind scanning produced very few good points,
even in simple CP conserving models.
For example,
Refs.~\cite{Das:2019yad,Boto:2021qgu} studied a $\mathbb{Z}_3$-symmetric Type Z 3HDM,
finding that simulations had to start close to the alignment limit
\cite{Craig:2013hca,Carena:2013ooa,Das:2019yad,Pilaftsis:2016erj},
where $h_{125}$ has properties very close to the SM Higgs boson.
Even worse performances are found in CP violating models.
For example, using the C3HDM to search blindly for points surviving all theoretical
bounds and all current experimental data found less than 1 valid point
per $10^{13}$ points sampled.
The solution adopted in Ref.~\cite{Boto:2024jgj} was to start close to the
CP-conserving limit and, within it, close to the alignment limit, slowly increasing
complex parameters.
This  is what we mean by an informed traditional simulation technique.

Of course, one can fear that starting searches around very specific points
in parameter space can lead to a skewed perception of a model's capabilities.
One may be mislead about a model's true potential, or, worse, even exclude a perfectly
viable model.
For example,
using a simulation starting from the real, aligned 3HDM,
Ref.~\cite{Boto:2024jgj} seemed to find that $c^o_{tt} \sim 0$ in the Type Z C3HDM.
This was disproved in Ref.~\cite{deSouza:2025bpl}, where Machine Learning  (ML)
techniques~\footnote{Various
methods have been devised to effectively
explore the parameter space of new Physics models when training datasets
are accessible
\cite{Caron:2019xkx,Hollingsworth:2021sii,Goodsell:2022beo,Diaz:2024yfu,Batra:2025amk,Hammad:2025wst}.
For a broader overview of Machine Learning
applications in particle physics, refer to the detailed reviews in
\cite{Feickert:2021ajf,Plehn:2022ftl}.
}
were used to survey fully the parameter space.
We looked into an Evolutionary Strategy Algorithm,
that does \textit{not} rely on valid training data,
finding good points of the model \cite{deSouza:2022uhk}.
An anomaly detection method for Novelty Reward \cite{Romao:2024gjx}
is incorporated to enhance exploration of the parameter space and, crucially,
the physical implications of the model.
Such methods have also been applied recently in order to explore
models with solutions to the DM problem \cite{deSouza:2025uxb,Boto:2025mmn}.

The existence of large CP-odd $h_{125}$-fermion couplings in the C3HDM but not in the
C2HDM provides one further indication that models with two Higgs doublets
are not the best suited to explore fully the possibilities available with NHDM.
This is already known from other sources.
For example,
in the 2HDM,
the existence of a normal vacuum with a massless photon ensures that
there will not exist a lower lying vacuum with a massive photon \cite{Ferreira:2004yd}.
This is no longer the case when there are three or more Higgs doublets \cite{Barroso:2006pa}.
In addition,
2HDM have only one charged scalar, which could lead to large
contributions to $b \rightarrow s \gamma$.
In some cases, this places a strong lower bound on the 
mass of the charged scalar \cite{Misiak:2017bgg};
see \cite{Misiak2025} for a recent update.
In contrast, in the 3HDM,
one can pit the contribution of one charged scalar against
that of the second charged scalar,
and one charged scalar mass as low as $200\textrm{GeV}$ is then possible
\cite{Hewett:1994bd,Akeroyd:2016ssd,Akeroyd:2020nfj,Boto:2021qgu}.

But, Multi Higgs models also bring new challenges.
In their most general form, they provoke natural flavour changing coupling (FCNC)
which are strongly constrained by precise measurements on neutral meson-antimeson systems,
such a $B - \bar{B}$.
This can be solved in the 2HDM by imposing a $\mathbb{Z}_2$
discrete symmetry \cite{Glashow:1976nt,Paschos:1976ay};
a mechanism known as natural flavour conservation (NFC).
The end result is that each charged fermion sector couples to a single
scalar field. In the 2HDM, there are four such possibilities.
In the 3HDM, Type Z becomes a possibility.
The renormalization group stability of the five distinct Types of Yukawa
couplings in models with more than two doublets
was discussed in \cite{Ferreira:2010xe} and denoted in \cite{Yagyu:2016whx} 
by Types I, II, X (also known as lepton-specific), Y (flipped), and Z (democratic),
according to 
\begin{eqnarray}
\textrm{Type I:}
&&
\Phi_u=\Phi_d=\Phi_e\, ,
\nonumber\\
\textrm{Type II:}
&&
\Phi_u \neq \Phi_d=\Phi_e\, ,
\nonumber\\
\textrm{Type X:}
&&
\Phi_u=\Phi_d \neq \Phi_e\, ,
\nonumber\\
\textrm{Type Y:}
&&
\Phi_u=\Phi_e \neq \Phi_d\, ,
\nonumber\\
\textrm{Type Z:}
&&
\Phi_u\neq \Phi_d;\ \Phi_d\neq \Phi_e,\ \Phi_e\neq \Phi_u\, .
\label{types}
\end{eqnarray}

The C3HDM model has five neutral scalars (typically with both CP-even and CP-odd components),
which we order by their mass,
with $h_1$ the lightest and $h_5$ the heaviest.
Having determined the possible model Types, one must still indicate which
of the 5 neutral scalars corresponds
to the observed Higgs scalar $h_{125}$.
Each identification of the Higgs scalar with one of the $h_i$ corresponds
to one possible ordering of our model, leading to five distinct scenarios.
To be specific, the $5$ orderings are:
\be
\textrm{h1:}\ \ h_1=h_{125}\, , \ \ \ \
\textrm{h2:}\ \ h_2=h_{125}\, , \ \ \ \ 
\textrm{h3:}\ \ h_3=h_{125}\, , \ \ \ \
\textrm{h4:}\ \ h_4=h_{125}\, , \ \ \ \
\textrm{h5:}\ \ h_5=h_{125}\, .
\label{ordering}
\ee
In the first case, the $125\,\textrm{GeV}$ Higgs found at LHC is the lightest neutral scalar in the theory;
in the other four cases, there are neutral scalars below $125\,\textrm{GeV}$ 
yet to be discovered.

Some versions of these models have been the subject of previous studies.
We list some example from
the literature on 3HDM with natural flavor conservation in
Table~\ref{tab:summary_literature}.
\begin{table}[h]
  \centering
  \begin{tabular}{|l|c|c|c|c|c|}\hline
    Type   & I & II& X & Y & Z\\ \hline
$h_1=h_{125}$&  \cite{Boto:2024tzp,Aranda:2019vda,Kuncinas:2024zjq,Keus:2014jha,Keus:2014isa,Cordero-Cid:2016krd,DuttaBanik:2025fce}
& \cite{DuttaBanik:2025fce,Merchand:2019bod}
& -
& -
& \cite{Boto:2021qgu,Chakraborti:2021bpy,Boto:2022uwv,Das:2022gbm,Boto:2023nyi,Boto:2024jgj,Romao:2024gjx,deSouza:2025bpl,Coleppa:2025qst,Batra:2025amk}
\\\hline
$h_2=h_{125}$ & \cite{Boto:2024tzp,Aranda:2019vda,Hernandez-Otero:2022dxd,Kuncinas:2024zjq}
& \cite{Merchand:2019bod}
& -
& -
&  \cite{Batra:2025amk,Plantey:2022jdg,Ogreid:2024qfw}
\\\hline
$h_3=h_{125}$ & \cite{Boto:2024tzp,Cordero-Cid:2018man,Aranda:2019vda,Hernandez-Otero:2022dxd,Cordero-Cid:2020yba,Kuncinas:2024zjq}
& \cite{Merchand:2019bod}
& -
& -
& \cite{Batra:2025amk,Plantey:2022jdg,Ogreid:2024qfw}
\\\hline
$h_4=h_{125}$
& \cite{Boto:2024tzp,Cordero-Cid:2018man,Aranda:2019vda,Hernandez-Otero:2022dxd,Cordero-Cid:2020yba,Kuncinas:2024zjq}
& \cite{Merchand:2019bod}
& -
&  -
& \cite{Batra:2025amk}
\\\hline
$h_5=h_{125}$
& \cite{Boto:2024tzp,Cordero-Cid:2018man,Aranda:2019vda,Hernandez-Otero:2022dxd,Cordero-Cid:2020yba,Kuncinas:2024zjq,Dey:2023exa}
& -
& -
& -
& -
\\\hline
  \end{tabular}
  \caption{Literature on 3HDM with Natural Flavor Conservation. \cite{Cordero-Cid:2016krd,Cordero-Cid:2018man,Boto:2024jgj,deSouza:2025bpl,Cordero-Cid:2020yba,Dey:2024epo}
have explicit CPV in the scalar sector.
\cite{Plantey:2022jdg,Ogreid:2024qfw} have spontaneous CPV.
\cite{Boto:2024tzp,Cordero-Cid:2018man,Aranda:2019vda,Keus:2014jha,Keus:2014isa,Cordero-Cid:2016krd,DuttaBanik:2025fce,Merchand:2019bod,Hernandez-Otero:2022dxd,Cordero-Cid:2020yba,Dey:2024epo,Kuncinas:2024zjq}
have inert doublets.
}
  \label{tab:summary_literature}
\end{table}

Our objective is to explore the possibility of large CP-odd $h_{125}$ couplings to fermions
across all Types and all possible mass orderings; twenty five cases in all.
In order to explore fully the models capabilities,
we utilize an evolutionary strategy algorithm with novelty reward.
This complete survey of large CP-odd components in the C3HDM will provide
examples for additional exploration and can help in
motivating further experimental searches.

We describe the scalar potential, Yukawa couplings and interaction with
gauge bosons of our model in Secs.~\ref{sec:pot}, \ref{sec:Yuk}, and \ref{sec:GaB},
respectively.
Sec.~\ref{sec:constraints} describes the full set of theoretical and experimental
constraints that our model is subjected to,
while Sec.~\ref{sec:scan} describes the scanning procedure utilized,
involving a Machine Learning black box optimization algorithm.
Our results for the bottom and/or tau are presented in Sec.~\ref{sec:results},
complemented by results for the top in appendix~\ref{app:top}  and some benchmark points in appendix~\ref{app:BPs}.
Subsection~\ref{subsec:scalarornot} is devoted to the peculiar situation that
$h_{125}$ may couple as a pure scalar to some fermions,
while it couples as a pure pseudoscalar to other fermions
We draw our conclusions in Sec.~\ref{sec:concl}.

\section{\label{sec:pot}The scalar Potential}

The general scalar potential for the C3HDM with a softly broken
$\mathbb{Z}_2\times\mathbb{Z}_2$ symmetry is

\begin{equation}
    V = V_2+V_4 = \mu_{ij}(\Phi_i^{\dagger}\Phi_j)
+ z_{ijkl}(\Phi_i^{\dagger}\Phi_j)(\Phi_k^{\dagger}\Phi_l), \label{Vpot}
\end{equation}
\
with
\
\begin{equation}
\begin{split}
        V_2=\,& \mu_{11}(\Phi_1^{\dagger}\Phi_1) + \mu_{22}(\Phi_2^{\dagger}\Phi_2)
+ \mu_{33}(\Phi_3^{\dagger}\Phi_3) +
\left[
\left(
\mu_{12}(\Phi_1^{\dagger}\Phi_2)
+ \mu_{13}(\Phi_1^{\dagger}\Phi_3) + \mu_{23}(\Phi_2^{\dagger}\Phi_3)
\right)
+ h.c.
\right], \\
		V_4 =\,& V_{RI} + V_{Z_2\times Z_2} \;,\\
		V_{RI} =\,& \lambda_1 (\Phi_1^\dag \Phi_1)^2
+ \lambda_2 (\Phi_2^\dag \Phi_2)^2 + \lambda_3 (\Phi_3^\dag \Phi_3)^2 
+ \lambda_4 (\Phi_1^\dag \Phi_1)(\Phi_2^\dag \Phi_2)
+ \lambda_5 (\Phi_1^\dag \Phi_1)(\Phi_3^\dag \Phi_3)
\\
&+ \lambda_6 (\Phi_2^\dag \Phi_2)(\Phi_3^\dag \Phi_3) 
+ \lambda_7 (\Phi_1^\dag \Phi_2)(\Phi_2^\dag \Phi_1)
+ \lambda_8 (\Phi_1^\dag \Phi_3)(\Phi_3^\dag \Phi_1)
+ \lambda_9 (\Phi_2^\dag \Phi_3)(\Phi_3^\dag \Phi_2) \;, \\
		V_{Z_2\times Z_2} =\,
& \lambda_{10} (\Phi_1^\dag \Phi_2)^2 + \lambda_{11} (\Phi_1^\dag \Phi_3)^2
+ \lambda_{12} (\Phi_2^\dag \Phi_3)^2 + h.c.
		\;, 
\end{split} \label{Vlong}
\end{equation}
where $h.c.$ stands for hermitian conjugation,
the complex $\mu_{12}$, $\mu_{13}$, {$\mu_{23}$} softly break
the $\mathbb{Z}_2\times \mathbb{Z}_2$ symmetry,
while the other $\mu_{11}$, $\mu_{22}$, $\mu_{33}$ are real.
Moreover, $\lambda_{10-12}$ are generally complex, and all the others
$\lambda$'s are real.
The fourth order potential, $V_4$, is separated into two parts:
$V_{RI}$, which is invariant for an independent rephasing of each doublet,
and $V_{Z_2\times Z_2}$, composed of terms allowed by the
$Z_2\times Z_2$ symmetry but not invariant under independent rephasings.

For the treatment of the potential, we follow closely the procedure
of \cite{Boto:2024jgj}, to which we refer for a more detailed description.
We start by parameterising the doublets  after 
Spontaneous Symmetry Breaking (SSB) as
\begin{equation}
	\Phi_i = 
	\begin{pmatrix}
		w_i^+ \\ (v_i + x_i + i\ z_i) /\sqrt{2}
	\end{pmatrix} 
	=
	\begin{pmatrix}
		w_i^+ \\ (v_i + \varphi_i) /\sqrt{2}
	\end{pmatrix} 
	\;,
\end{equation}
where the vacuum expectation values (vev), $v_i$, although generally complex,
can be made real by a change of basis that relocates those
phases to the potential parameters. 
To determine the relations between the vevs, masses and potential parameters,
we start by defining two auxiliary hermitian matrices and one symmetric matrix,
given respectively by
\begin{eqnarray}
A_{ij} &=&
\mu_{ij} + z_{ijkl} v_k^* v_l \, ,
\nonumber\\
B_{ij} &=&
z_{iklj} v_l^* v_k\, ,
\nonumber\\
C_{ij} &=&
z_{kilj} v_l^* v_k^* \, .
\end{eqnarray}
We compute the linear and quadratic parts of the potential:
\begin{equation}
\begin{split}
    V^{(1)} =\ &\boldsymbol{x}^T\text{Re}(A\boldsymbol{v})-i\boldsymbol{z}^T\text{Im}(Av) \, ,\\
    V^{(2)} =\ & V^{(2)}_{ch} + V^{(2)}_{n} =(\boldsymbol{w}^+)^\dagger M^2_{ch} \boldsymbol{w}^+ +  
		\frac{1}{2} \begin{pmatrix}
			\boldsymbol{x}^T & \boldsymbol{z}^T
		\end{pmatrix}
		M^2_n
		\begin{pmatrix}
			\boldsymbol{x} \\ \boldsymbol{z}
		\end{pmatrix}\; ,
\end{split}
\end{equation}
where the mass matrices are:
\begin{equation}\begin{split}
M^2_n&=\begin{pmatrix}
			M^2_x & M^2_{xz} \\
			(M^2_{xz})^T & M^2_z 
		\end{pmatrix}  \;, \\
		M^2_{ch} &= A \;, \\
		M^2_x &= \Re(A+B + C) \;, \\
            M^2_z &= \Re(A+B - C) \;, \\
		M^2_{xz} &= -\Im(A+B+C) \; .
\end{split}\end{equation}
The relations between the vev components and the potential parameters
are obtained through the stationary condition: $V^{(1)}=0\implies A\boldsymbol{v}=0$.
Those take the form:
\small
\begin{equation}\begin{split}\label{STAT}
		\mu_{11} v_1 = -\Re(\mu_{12}) v_2 -\Re(\mu_{13}) v_3 -
&v_1 \left[ \lambda_1 v_1^2 + \left(\Re(\lambda_{10}) + \frac{1}{2}\lambda_4
+ \frac{1}{2}\lambda_7\right) v_2^2 + \left(\Re(\lambda_{11})
+ \frac{1}{2} \lambda_5 + \frac{1}{2}\lambda_8\right) v_3^2 \right] 
		\;, \\*[1mm]
		\mu_{22} v_2 =  -\Re(\mu_{12}) v_1 -\Re(\mu_{23}) v_3 -
&v_2 \left[ \lambda_2 v_2^2 + \left(\Re(\lambda_{10}) + \frac{1}{2}\lambda_4
+ \frac{1}{2}\lambda_7\right) v_1^2 + \left(\Re(\lambda_{12})
+ \frac{1}{2}\lambda_6 + \frac{1}{2}\lambda_9\right) v_3^2 \right] 
		\;, \\*[1mm]
		\mu_{33} v_3 =  -\Re(\mu_{13}) v_1 -\Re(\mu_{23}) v_2 - 
&v_3 \left[ \lambda_3 v_3^2 + \left(\Re(\lambda_{11}) + \frac{1}{2} \lambda_5
+ \frac{1}{2}\lambda_8\right) v_1^2 + \left(\Re(\lambda_{12}) + \frac{1}{2}\lambda_6
+ \frac{1}{2}\lambda_9\right) v_2^2\right] 
		\;, \\*[1mm]
		\Im(\mu_{13}) v_3
&=  -v_1\left[ \Im(\lambda_{10}) v_2^2
+ \Im(\lambda_{11}) v_3^2 \right] - \Im(\mu_{12}) v_2 
		\;, \\*[1mm]
		\Im(\mu_{23}) v_3 
&= v_2\left[ \Im(\lambda_{10}) v_1^2
- \Im(\lambda_{12})v_3^2 \right] + \Im(\mu_{12}) v_1
		\;.
	\end{split}
\end{equation}
\normalsize
For a charge conserving vev, these conditions, when applied to the mass matrices,
lead to two massless scalars: one charged $G^{+}=\frac{v_i}{v}\omega_i^{+}$
and one neutral $G^{0}=\frac{v_i}{v}z_i$. These correspond to the would-be
Goldstone bosons that, in unitary gauge,
become the longitudinal components of the weak force bosons.
The five stationary conditions are applied to the original 24 parameters
in the scalar potential of \eq{Vlong}, plus the three real vevs.
After fixing two parameters to be $v$ and the mass of $125\textrm{GeV}$,
the model has 20 free parameters to sample with.

We now shift our focus to obtaining the scalar masses from the mass matrices.
First, we deal with the massless states, which we can factor out as
they are aligned with the vev direction:
\begin{equation}
    \begin{pmatrix}
			v_1 \\
            v_2 \\
            v_3
	\end{pmatrix} =
    v\begin{pmatrix}
        c_{\beta_2}c_{\beta_1} \\
        c_{\beta_2}s_{\beta_1} \\
        s_{\beta_2} 
    \end{pmatrix}
    .
\end{equation}
We do this through a rotation to the Higgs basis \cite{Botella:1994cs},
in which the vev and the massless states are found only in the first
doublet \cite{Georgi:1978ri,Donoghue:1978cj}.
However, in order to simplify the equations, we only rotate the charged
and CP-odd scalars (instead of the doublets as a whole),
as the two massless states
are combinations of each of these types only.
The rotation matrix and the new mass matrices are as follows:
\begin{equation}
	R_H = R_{13}(\beta_2) R_{12}(\beta_1) = 
	\begin{pmatrix}
		c_{\beta_2} & 0 & s_{\beta_2} \\
		0 & 1 & 0 \\
		-s_{\beta_2} & 0 & c_{\beta_2} 
	\end{pmatrix}
	\begin{pmatrix}
		c_{\beta_1} & s_{\beta_1} & 0 \\
		-s_{\beta_1} & c_{\beta_1} & 0 \\
		0 & 0 & 1 
	\end{pmatrix}
	=
	\begin{pmatrix}
		c_{\beta_2}c_{\beta_1} & c_{\beta_2}s_{\beta_1} & s_{\beta_2} \\
		-s_{\beta_1} & c_{\beta_1} & 0 \\
		-s_{\beta_2}c_{\beta_1} & -s_{\beta_2}s_{\beta_1} & c_{\beta_2} 
	\end{pmatrix}\, ,
\end{equation}
\begin{equation}\label{e:field_transf}
	\begin{pmatrix}
		x \\ z^\prime
	\end{pmatrix}
	= \begin{pmatrix}
		\one & 0 \\
		0 & R_H
	\end{pmatrix}
	\begin{pmatrix}
		x \\ z
	\end{pmatrix}
	\;,
	\qquad \qquad
	w^{+\prime} = R_H w^+
	\; ,
\end{equation}
\begin{equation}
 R_H M_{ch}^2 R_H^T =\begin{pmatrix}
     0 & 0 & 0  \\
    0 &\multicolumn{2}{c}{\multirow{2}{*}{ $ M_{ch}^{2\prime}$}}\\ 
    0  &
    \end{pmatrix}\;, 
\qquad  \quad  
\begin{pmatrix}
		\one & 0 \\
		0 & R_H
	\end{pmatrix}M_{n}^2\begin{pmatrix}
		\one & 0 \\
		0 & R_H^T
	\end{pmatrix}=\begin{pmatrix}
\multicolumn{3}{c}{\multirow{3}{*}{ $ M_{x}^{2}$}} 
& 0 & \multicolumn{2}{c}{\multirow{3}{*}{ $ M_{xz}^{2\prime}$}}\\ 
      &  &  & 0 &  &   \\
           &  &  & 0 &  &   \\
     0 & 0 & 0 & 0 & 0 & 0 \\
    \multicolumn{3}{c}{\multirow{2}{*}{ $ M_{xz}^{2\prime T}$}}
& 0 & \multicolumn{2}{c}{\multirow{2}{*}{ $ M_z^{2\prime}$}}\\ 
     &  &  & 0 &  &   
    \end{pmatrix}
		\;.
\end{equation}
Finally, with the massless states factored out and having identified
the reduced mass matrices, denoted by primes,
we can now perform general unitary transformations to arrive at the
other 2 charged and 5 neutral mass eigenstates, through
\begin{equation}\label{mass_eigenstates}
	\Big(H_1^+ \; H_2^+\Big)^T = W\;\Big(w_2^{+\prime}\; w_3^{+\prime}\Big)^T  \;, \qquad \quad \Big(h_1\;h_2\;h_3\;h_4\;h_5\Big)^T = R 
	\;\Big(x_1\;x_2\;x_3\;z_2^\prime\; z_3^\prime\Big)^T \;,
\end{equation}
with $W$ and $R$ such that
\begin{equation} \label{e:mass_equations}
	M_{ch}^{2\prime} = W^\dag 
	\begin{pmatrix}
		m_{H_1^\pm}^2 & 0 \\ 0 & m_{H_2^\pm}^2
	\end{pmatrix}
	W
	\;,
	\qquad \quad
	M_{n}^{2\prime} = R^T 
	\begin{pmatrix}
		m_{h_1}^2 & 0 & 0 & 0 & 0 \\ 
		0 & m_{h_2}^2 & 0 & 0 & 0 \\ 
		0 & 0 & m_{h_3}^2 & 0 & 0 \\ 
		0 & 0 & 0 & m_{h_4}^2 & 0 \\ 
		0 & 0 & 0 & 0 & m_{h_5}^2 
	\end{pmatrix}
	R
	\;,
\end{equation}
where $m_{H^{\pm}_{i}}^2$ and $m_{h_i}^2$ are the squared masses of
the charged and neutral scalars, respectively, ordered in increasing mass. 
Although the unitary matrix $W$ can be simply parameterised by 2 angles,
$\theta$ and $\varphi$, the orthogonal matrix $R$ is the product of 10
rotation matrices parameterised by 10 angles,
labeled $\alpha_{ij}$, with $j>i$.

Equation (\ref{e:mass_equations}) gives us 19 linear real equations between
the potential parameters, the squared masses and the $R$ matrix entries,
or more precisely, the 10 rotation angles. 
Sixteen of these equations let us exchange the potential parameters by the
squared masses and the rotation angles,
while the three heavier neutral scalar masses
$m_{h_{3,4,5}}$ are derived from the remaining 3 equations.

The only thing left to state is which of the 5 neutral scalars corresponds
to the observed Higgs scalar with a mass of $125\,\textrm{GeV}$ ($h_{125}$).
We study all possible orderings, according to the nomenclature in 
Eq.~\eqref{ordering} and following the mass ordering of Eq.~\eqref{e:mass_equations}.

\section{\label{sec:Yuk}The Yukawa Lagrangian}

Due to the $Z_2\times Z_2$ symmetry, each fermion type only couples to one of the doublets.
To remain as general as possible,
we name the coupling doublets as $\Phi_u$, $\Phi_d$ and $\Phi_\ell$,
where the subscript indicates which charge fermion type the scalar couples to.
The five possible Yukawa Types are listed in Eq.~\ref{types}.
With this, the Yukawa Lagrangian reads:
\begin{equation}\label{LY}
- \mathcal{L}_\text{Yukawa} = \overline{Q}_L \Gamma \Phi_d n_R
+\overline{Q}_L \Delta \tilde{\Phi}_u p_R
+\overline{L}_L Y \Phi_\ell \ell_R + h.c. 
\;,
\end{equation}
where $Q_L = (p_L \; n_L)^T$, $L_L = (\nu_L \; \ell_L)^T$,
while $n_R$, $p_R$, and $\ell_R$ are left-handed doublets and
right-handed singlets of the different fermions on the weak basis.
Analogous to the SM, the SSB mechanism results in the scalar fields having
vevs that lead to mass terms for the fermions. Considering massless neutrinos,
the charged lepton basis can be chosen such that
\begin{equation}
M_{\ell}= \frac{v_\ell}{\sqrt{2}}Y
\end{equation}
is already diagonal, $\text{diag}(m_e, m_\mu, m_\tau)$. 
To change from the flavour to the diagonal basis,
the quark fields require unitary transformations of the form,
\begin{equation}
 p_L=U_L^p u_L , \quad p_R=U_R^p u_R\, ,\quad n_L=U_L^n d_L\, , \quad n_R=U_R^n d_R\,\, ,
\end{equation}
that diagonalize the mass terms,
\begin{equation}
   (U_L^n)^\dagger \frac{v_d}{\sqrt{2}}\Gamma\, U_R^n \equiv   D_d\, ,
\quad (U_L^p)^\dagger \frac{v_u}{\sqrt{2}}\Delta\, U_R^p\equiv  D_u,
\end{equation}
leading to the fermion masses,
$D_u=\text{diag}(m_u,m_c,m_t)$,
and $D_d=\text{diag}(m_d,m_s,m_b)$,
and the CKM \cite{bib:CKM-Cabibbo,Kobayashi:1973fv}
matrix $V_{\text{CKM}}=(U^p_L)^\dagger U^n_L$.
With multiple doublets,
the fermion-scalar interactions have multiple possibilities based on the Type of assignments.
These can be deduced from the  Yukawa Lagrangian,
after writing the fermion fields in the mass basis as
\begin{equation}
	\begin{split}
		- \mathcal{L}_{\Phi ff} = 
		 \left(\overline{u}_L V_{\text{CKM}} \;\;\; \overline{d}_L \right)
\frac{\sqrt{2} D_d}{v_d} \begin{pmatrix}
			w_d^+ \\ (x_d+iz_d)/\sqrt{2}
		\end{pmatrix} d_R
		+ \left(\overline{u}_L \;\;\; \overline{d}_L V_{\text{CKM}}^\dagger \right)
&\frac{\sqrt{2} D_u}{v_u} \begin{pmatrix}
			(x_u-iz_u)/\sqrt{2} \\ -w_u^-
		\end{pmatrix} u_R 
		\\*[1mm]
		+\, (\overline{\nu}_L \; \overline{\ell}_L)
\frac{\sqrt{2} D_\ell}{v_\ell} \begin{pmatrix}
			w_\ell^+ \\ (x_\ell+i z_\ell)/\sqrt{2}
		\end{pmatrix}& \ell_R
		+ h.c. \, .
	\end{split}
\end{equation}
As we are interested in the couplings to the observed Higgs boson,
we highlight the neutral scalar interactions:
\begin{equation} \label{Lh0ff}
	-\mathcal{L}_{\xi ff} = 
	\frac{(m_{\ell})_i}{v_\ell} \overline{\ell}_i (x_\ell + i \gamma_5 z_\ell) \ell_i 
	+\frac{(m_{d})_i}{v_d} \overline{d}_i (x_d + i \gamma_5 z_d) d_i 
	+\frac{(m_{u})_i}{v_u} \overline{u}_i (x_u - i \gamma_5 z_u) u_i 
	\; .
\end{equation}
The current Lagrangian is written in terms of the original scalar states.
Thus, to extract the wanted couplings, we need to rewrite the Lagrangian
with the neutral scalars in the mass basis.
Following the scalar potential treatment that led to \eqs{e:field_transf}{mass_eigenstates},
the mass states are obtained from the symmetry basis through a rotation matrix $Q$,
which is a combination of the
rotation matrix that leads to the Higgs basis and the $R$ matrix,
\begin{equation}\label{matrixQ}
    \begin{split}
    \begin{pmatrix}
			\xi_1 \\  \xi_2 \\ \xi_3 \\ \xi_4 \\ \xi_5 \\ \xi_6
		\end{pmatrix}
        \equiv
		\begin{pmatrix}
			G^0 \\ h_1 \\ h_2 \\ h_3 \\ h_4 \\ h_5
		\end{pmatrix}
        		&=
		Q
		\begin{pmatrix}
			x_1 \\ x_2 \\ x_3 \\ z_1 \\ z_2 \\ z_3
		\end{pmatrix}
\\
	&= 		\left(\begin{matrix}0 & 0 & 0 & c_{\beta_1} c_{\beta_2}
& c_{\beta_2} s_{\beta_1} & s_{\beta_2}
\\
R_{11} & R_{12} & R_{13} & - R_{14} s_{\beta_1} - R_{15} c_{\beta_1} s_{\beta_2}
& R_{14} c_{\beta_1} - R_{15} s_{\beta_1} s_{\beta_2} & R_{15} c_{\beta_2}
\\
R_{21} & R_{22} & R_{23} & - R_{24} s_{\beta_1} - R_{25} c_{\beta_1} s_{\beta_2}
& R_{24} c_{\beta_1} - R_{25} s_{\beta_1} s_{\beta_2} & R_{25} c_{\beta_2}
\\
R_{31} & R_{32} & R_{33} & - R_{34} s_{\beta_1} - R_{35} c_{\beta_1} s_{\beta_2}
& R_{34} c_{\beta_1} - R_{35} s_{\beta_1} s_{\beta_2} & R_{35} c_{\beta_2}
\\
R_{41} & R_{42} & R_{43} & - R_{44} s_{\beta_1} - R_{45} c_{\beta_1} s_{\beta_2}
& R_{44} c_{\beta_1} - R_{45} s_{\beta_1} s_{\beta_2} & R_{45} c_{\beta_2}
\\
R_{51} & R_{52} & R_{53} & - R_{54} s_{\beta_1} - R_{55} c_{\beta_1} s_{\beta_2}
& R_{54} c_{\beta_1} - R_{55} s_{\beta_1} s_{\beta_2} & R_{55} c_{\beta_2}
\end{matrix}\right)
		\begin{pmatrix}
			x_1 \\ x_2 \\ x_3 \\ z_1 \\ z_2 \\ z_3
		\end{pmatrix} ,
	\end{split}
\end{equation}
identifying the would-be Goldstone in the first position, $\xi_1=G^0$.
Eq.~\eqref{matrixQ} can equivalently be written as,
\begin{equation}
	x_i = Q^T_{ij}\; \xi_j = Q_{ji} \;\xi_j
\quad \text{and}
\quad z_i = Q^T_{3+i,j}\; \xi_j = Q_{j,3+i} \;\xi_j
	\;,
\end{equation}
with $i$ running from 1 to 3 and $j$ from 1 to 6.
Substituting in the Lagrangian, we get
\begin{equation}\begin{split}
		-\mathcal{L}_{\xi ff} = 
		\sum_f \sum_{j=1}^{6}  \frac{m_f}{v} \overline{f}
\frac{v}{v_f}\left( Q_{jf} \pm i \gamma_5 Q_{j,3+f}\right) f\; \xi_j 
		=
		\sum_f \sum_{j=1}^{6}  \frac{m_f}{v} \overline{f}
\left( c^e_{\xi_j ff} + i\gamma_5 c^o_{\xi_j ff} \right) f\; \xi_j 
		\;,
\end{split}\end{equation}
where we have defined 
\begin{equation}\label{h0ff_couplings}
	\;\; c^e_{\xi_j ff} + i\gamma_5 c^o_{\xi_j ff} =
\frac{v}{v_f}\left( Q_{jf} \pm i \gamma_5 Q_{j,3+f}\right)
	\;,
\end{equation}
and the minus sign appears only for the up-type quarks,
as they use the conjugate doublet in the Yukawa Lagrangian.

\subsection{Higgs fermion couplings}

Until now, the treatment performed has been independent of the ordering
chosen in Eq.~\eqref{ordering}
and the Type chosen
in Eq.~\eqref{types}.
These two choices are reflected in the couplings between
the fermions and scalars, especially with $h_{125}$.
Starting with the ordering, it will determine which of the $h_i$
is the Higgs one, hence it will choose the $\xi_j$
in equation (\ref{h0ff_couplings}) that corresponds to the $h_{125}$,
thus fixing the $j$ index in that equation.
On the other hand, the chosen Type determines to what $\Phi_i$
each one of $\Phi_u$, $\Phi_d$, and $\Phi_\ell$ corresponds to.
This assigns one of $x_{i}$ to each of $x_{u/d/\ell}$  in equation
\eqref{Lh0ff} and similarly to the $z$'s.
Hence, this translates to specifying a number from 1 to 3
to each $f$ in equation \eqref{h0ff_couplings}.
In other words, when choosing the fermion on the left side,
the choice of Type determines which number $f$ corresponds to on the right side.
This results in the following couplings:
\begin{equation}
	c^e_{h_{125}ff} =  
	\frac{R_{i1}}{c_{\beta_2} c_{\beta_1}} \;,\;
	\frac{R_{i2}}{c_{\beta_2} s_{\beta_1}} \;,\;
	\frac{R_{i3}}{s_{\beta_2} } \;,\;
	\qquad \text{for} \quad f=1,2,3 \;,
\end{equation}
and
\begin{equation}
	c^o_{h_{125}ff} = 
	\pm\frac{-R_{i4}s_{\beta_1}-R_{i5}c_{\beta_1}s_{\beta_2}}{c_{\beta_2} c_{\beta_1}} \;, \;
	\pm\frac{R_{i4}c_{\beta_1}-R_{i5}s_{\beta_1}s_{\beta_2}}{c_{\beta_2} s_{\beta_1}} \;,\;
	\pm  \frac{R_{i5} c_{\beta_2}}{s_{\beta_2}}
	\qquad \text{for} \quad f=1,2,3 \; ,
\end{equation}
where $i=j+1$ indicates the chosen ordering and $f$ corresponds to
the index assigned to fermions of a given electric charge by the chosen Type. We verified the couplings for each Type with Feynmaster
\cite{Fontes:2021iue,Fontes:2025svw} and list them in Table~\ref{table:I}:
\begin{table}[h!]
\begin{tabular}{r @{\hskip 0.5cm} c c @{\hskip 0.5cm} c c @{\hskip 0.5cm} c c}
\hline
& \multicolumn{2}{c}{leptons} & \multicolumn{2}{c}{$d$-type} & \multicolumn{2}{c}{$u$-type} \\
& $c^e$ & $c^o$ & $c^e$ & $c^o$ & $c^e$ & $c^o$ \\
\hline \\[-0.3cm]
Type I & $\frac{R_{i3}}{s_{\beta_2}}$ & $\frac{R_{i5}c_{\beta_2}}{s_{\beta_2}}$ &
$\frac{R_{i3}}{s_{\beta_2}}$ & $\frac{R_{i5}c_{\beta_2}}{s_{\beta_2}}$ &
$\frac{R_{i3}}{s_{\beta_2}}$ & $-\frac{R_{i5}c_{\beta_2}}{s_{\beta_2}}$ \\[0.2cm]
Type II & $\frac{R_{i2}}{c_{\beta_2}s_{\beta_1}}$ & $\frac{R_{i4}c_{\beta_1}-R_{i5}s_{\beta_1}s_{\beta_2}}{c_{\beta_2}s_{\beta_1}}$ &
$\frac{R_{i2}}{c_{\beta_2}s_{\beta_1}}$
& $\frac{R_{i4}c_{\beta_1}-R_{i5}s_{\beta_1}s_{\beta_2}}{c_{\beta_2}s_{\beta_1}}$ &
$\frac{R_{i3}}{s_{\beta_2}}$ & $-\frac{R_{i5}c_{\beta_2}}{s_{\beta_2}}$ \\[0.2cm]
Type X & 
$\frac{R_{i2}}{c_{\beta_2}s_{\beta_1}}$
& $\frac{R_{i4}c_{\beta_1}-R_{i5}s_{\beta_1}s_{\beta_2}}{c_{\beta_2}s_{\beta_1}}$ &
$\frac{R_{i3}}{s_{\beta_2}}$ & $\frac{R_{i5}c_{\beta_2}}{s_{\beta_2}}$
& $\frac{R_{i3}}{s_{\beta_2}}$ & $-\frac{R_{i5}c_{\beta_2}}{s_{\beta_2}}$ \\[0.2cm]
Type Y & $\frac{R_{i3}}{s_{\beta_2}}$ & $\frac{R_{i5}c_{\beta_2}}{s_{\beta_2}}$
&
$\frac{R_{i2}}{c_{\beta_2}s_{\beta_1}}$
& $\frac{R_{i4}c_{\beta_1}-R_{i5}s_{\beta_1}s_{\beta_2}}{c_{\beta_2}s_{\beta_1}}$
& $\frac{R_{i3}}{s_{\beta_2}}$ & $-\frac{R_{i5}c_{\beta_2}}{s_{\beta_2}}$ \\[0.2cm]
Type Z & $\frac{R_{i1}}{c_{\beta_1}c_{\beta_2}}$ & $\frac{-R_{i4}s_{\beta_1}-R_{i5}c_{\beta_1}s_{\beta_2}}{c_{\beta_1}c_{\beta_2}}$
&
$\frac{R_{i2}}{c_{\beta_2}s_{\beta_1}}$
& $\frac{R_{i4}c_{\beta_1}-R_{i5}s_{\beta_1}s_{\beta_2}}{c_{\beta_2}s_{\beta_1}}$
&
$\frac{R_{i3}}{s_{\beta_2}}$ & $-\frac{R_{i5}c_{\beta_2}}{s_{\beta_2}}$ \\[0.2cm]
\hline
\end{tabular}
\caption{CP-even and CP-odd scalar-fermion couplings for each model Type
and fermions of a given electric charge.
The index $i$ reflects the chosen ordering.
\label{table:I}}
\end{table}

\section{\label{sec:GaB}Gauge Boson couplings}

To end the theoretical exposition of the model, we turn to the
couplings between the gauge bosons and the scalars,
especially to the Higgs boson.
For this, we take advantage of the fact that, in the Higgs Basis~\cite{Botella:1994cs},
the first doublet is solely responsible for the SSB~\cite{Georgi:1978ri,Donoghue:1978cj}.
Thus, only its CP-even scalar acquires a trilinear vertex of
the form $hVV$ with the Weak Bosons ($V=W,Z$). 

Recalling the relation
\begin{equation}
x_1'= (R_H)_{1i}x_i=(R_H)_{1i}Q_{ji}\xi_j\, ,
\end{equation}
the Lagrangian governing these interactions is,
\begin{equation}\label{e:kV}
	\begin{split}
		\mathcal{L} \supseteq &  
		\left(\frac{g^2 v}{2} W_\mu^+ W_\mu^- +  \frac{g^2 v}{4 c_W^2} Z_\mu^2 \right) x_1^\prime
=
		\left(\frac{g^2 v}{2} W_\mu^+ W_\mu^- + \frac{g^2 v}{4 c_W^2} Z_\mu^2 \right) 
		(R_H)_{1i} Q_{ji} \xi_j\, .
	\end{split}
\end{equation}
Let us take a particular ordering in Eq.~\eqref{ordering},
corresponding to a specific choice for $j$.
The piece of the Lagrangian referring to that specific $j$ can be written as
\be
\mathcal{L}^{(SM)}_{hVV} \kappa_V \, ,
\ee
where
\be
\kappa_V
= \left(\begin{pmatrix}
R_H & 0 \\ 0 & R_H
\end{pmatrix}
Q^T\right)_{12} = (R_H)_{1i} Q_{ji}\, .
\ee
Recall that,
in the last equation the index $j$ stands for the chosen ordering.

Notice that one can change the sign of any scalar field,
thus changing the sign of all its couplings to fermions and gauge bosons.
Thus, individually considered,
no one such sign has a physical significance.
However, relative signs do have physical significance.
For example,
the SM predicts $\textrm{sign}(k_V) c^e_{bb} = +1$.
In contrast, even in the real 2HDM there is the possibility that
$\textrm{sign}(k_V) c^e_{bb} = -1$;
this is dubbed the ``wrong-sign'' solution
\cite{Carmi:2012yp,Chiang:2013ixa,Ferreira:2014naa,Fontes:2014tga,Das:2022gbm}.

\section{\label{sec:constraints}Constraints}

The diagonalization procedure in \eq{e:mass_equations} leads to equations
that are solved for the heavier squared masses $m^2_{h_{3,4,5}}$,
which  must be restricted to all be positive and in the desired ordering,
identifying which $h_i$ is the $125\,\textrm{GeV}$ Higgs.
This is set as a
constraint that must be satisfied \textit{before}
the fitting procedure~\cite{deSouza:2025riy}.
During the simulation that follows,
all the remaining theoretical and experimental constraints
are treated in the same way.
This method enables independence from any predictive insight,
without attempting to identify before the fitting procedure which
particular observables will be easier to obey,
and which will turn out to be very difficult.

The constraints are:
boundedness from below \cite{Boto:2024jgj, Boto:2022uwv, Kannike_2012,Klimenko:1984qx};
perturbativity of the Yukawa couplings;
unitarity \cite{Bento_2022};
oblique parameters STU \cite{Grimus_2008, Baak2014};
$3\sigma$ experimental  $b \to s \gamma$ limit \cite{Florentino:2021ybj,Boto:2021qgu, Akeroyd:2020nfj};
the eEDM expressions in
\cite{Barr:1990vd,Yamanaka:2013pfn,Abe:2013qla,Inoue:2014nva,Altmannshofer:2020shb},
with experimental constraints  \cite{ACME:2018yjb,Roussy:2022cmp};
bounds of $|\theta_{\tau}| = |\arctan(c^o_{\tau\tau} / c^e_{\tau\tau})| < 34^\circ$ 
on CP-odd $H\tau\tau$ couplings \cite{CMS:2021sdq,ATLAS:2022akr}.
For the LHC signal strengths of the $125\textrm{GeV}$ Higgs we use
a very fast in-house code utilizing
Refs.\cite{Spira:1995mt,Frederix:2011zi,Broggio:2017oyu,Fontes:2021iue,Fontes:2025svw,Fontes:2014xva},
demanding that all signal strengths have a $2\sigma$ agreement with the
most recent ATLAS results \cite{ATLAS:2022vkf},
which are also consistent with CMS \cite{CMS:2022dwd}.
%This results in \textit{red} points, where there may be a large number of observables
%which lie at the edge of their allowed regions, yielding a large overall $\Delta \chi^2$.
%The points in \textit{green} combine points originally in red that are later found to
%also satisfy $\Delta \chi^2$ at $3\sigma$, calculated using \texttt{HiggsSignals-2}
%module in \texttt{HiggsTools-1.1.3}, with points generated with
%$\Delta \chi^2$ at $3\sigma$ as a constraint in the machine learning algorithm.
%This color code is applied for all Figures.

\section{\label{sec:scan}Scan Procedure}

The 20 free parameters, after setting the fixed inputs
$v = 246\,\text{GeV}$ and $m_{h_{125}} = 125\,\text{GeV}$, were sampled for similar  regions in each
model:~\footnote{Degenerate $125\,\text{GeV}$ scalars would require a
separate interpretation of the data. This has been avoided, since no new CP violating
features are foreseen.}
\begin{equation}
\begin{split}
&m_{h_i<h_{125}} \in [15.0,122.5] \textrm{GeV}\,;
\quad m_{h_i>h_{125}} \in [127.5,1000] \textrm{GeV}\,; 
\quad m_{H^\pm_k} \in [100,1000] \textrm{GeV}\,; 
\\*[3mm]
&\tan{\beta_1}, \tan{\beta_2} \in [0.3,10.0]\,;
\quad \Re(m_{12}^2),\Re(m_{13}^2),\Re(m_{23}^2)
\in [\pm 10^{-1},\pm 10^7]\, ; \\*[3mm]
&\theta, \phi, \alpha_{12}, \alpha_{13},
\alpha_{14}, \alpha_{15}, \alpha_{23}, \alpha_{24},
\alpha_{25}, \alpha_{34}, \alpha_{35}, \alpha_{45}
\in [-\pi,\pi]\,; 
\end{split}
\label{scaned_region}
\end{equation}
where $m_{h_i<h_{125}}$ ($m_{h_i>h_{125}}$) represent the masses of the neutral scalars
lighter (heavier) than the $h_{125}$, respectively.

The parameter sampling respecting Eq.~\eqref{scaned_region}
was performed with the Artificial Intelligence black box optimization approach first
presented in~\cite{deSouza:2022uhk},
and applied to the Type Z complex 3HDM in~\cite{deSouza:2025bpl}, for the choice $h_1=h_{125}$.
Alongside the theoretical and experimental constraints,
the Machine Learning (ML) algorithm considers a penalty system that pushes the algorithm
to explore different values of chosen parameters and/or observables.
This is known in general as \textit{novelty reward}.
Given a particular subset of parameters/observables that one applies novelty reward to, 
one says that one is performing a scan with \textit{focus}
on those parameters/observables.
Using the expressions in Table~\ref{table:I},
the chosen focus quantities for the Types I, II and X were the lepton couplings,
while for the Y and Z Types, the down-type couplings were chosen.
Furthermore, by considering additional cuts,
the algorithm allows us to concentrate on specific parameter and observable regions,
such as populating the wrong-sign, that would be otherwise difficult to sample.

Overall, this procedure was proven efficient in all of the different Types
and ordering combinations
(except for the cases in which it could not find any valid point).
Even in the failed scenarios, the algorithm was able
to indicate which constraints it struggled to satisfy, identifying
the reason for the model choice to become excluded.

\section{\label{sec:results}Results}

In the following subsections, we present our results for the combinations of five
Types of Yukawa
couplings and five possible choices of which of the neutral states
$h_i$ the $125\,\textrm{GeV}$ corresponds to.
The constraints imposed are identical for all of them,
and listed in Section~\ref{sec:constraints}. 
As mentioned,
the direct experimental constraints on CP-odd $h_{125}\tau\tau$ couplings
force
$|\theta_{\tau}| = |\arctan(c^o_{\tau\tau} / c^e_{\tau\tau})| < 34^\circ$
at $95\%$ CL \cite{CMS:2021sdq,ATLAS:2022akr}.
Similarly,
direct experimental constraints on CP-odd $h_{125}tt$ couplings
force
$|\theta_{t}| = |\arctan(c^o_{tt} / c^e_{tt})| < 43^\circ$ at $95\%$ CL \cite{ATLAS:2020ior}.
In contrast,
there is no direct experimental constraint on $c^o_{bb}$.
However,
in Type I,
the couplings of all charged fermions are equal.
Thus,
there,
$c^o_{bb}$ is limited by both the top and the tau direct constraints.
This situation is denoted by ``$t,\tau$'' in Table~\ref{tab:summary}.
Similarly,
in Type II and in Type X,
$c^o_{bb}= c^o_{\tau\tau}$ (denoted by ``$\tau$'')
and $c^o_{bb}= c^o_{tt}$ (denoted by ``$t$''),
respectively.
The hope for maximal $c^o_{bb}$ ($c^e_{bb}=0$) lies in Type Y and Type Z models.

In the complex two Higgs doublet model,
$c^o_{tt} \simeq 0$ for all four Types available in that model,
even if excluding direct experimental searches for CP-odd couplings.
Thus, Ref.~\cite{Biekotter:2024ykp} highlighted the importance of the $c^o_{\tau\tau}$
measurements in that case.
In contrast,
for the C3HDM,
$c^o_{tt}$ can be large.
In order to highlight the importance that the direct experimental constraints on
$c^o_{tt}$ have on these models,
we use the direct constraints on $c^o_{\tau\tau}$ in our simulation,
but do not impose the direct constraints on $c^o_{tt}$ a priori;
these will be shown by overlaying the current experimental lines from \cite{ATLAS:2020ior}.
We show the corresponding plots~\footnote{Except for Type I,
where all fermion couplings coincide.} in a dedicated
appendix~\ref{app:top}.

Since the signs of $c^{e}_f$ and $c^{o}_f$ have no absolute meaning
and are relative to the sign of
$k_V \equiv c(h_{125}VV)$, 
our plots are always shown with combinations of 
$\text{sgn}(k_V) c_f^o$ \textit{vs.}~$\text{sgn}(k_V) c_f^e$.
The overall situation is summarized in Table~\ref{tab:summary}.
\begin{table}[h]
  \centering
  \begin{tabular}{|l|c|c|c|c|c|}\hline
    Type   & I & II& X & Y & Z\\ \hline
$h_1=h_{125}$&$t,\tau$ &$\tau$ &$t$ & $\checkmark$ &  $\checkmark$\\\hline
$h_2=h_{125}$ &$t,\tau$ &$\tau$&$t$ &  $\checkmark$&  $\checkmark$\\\hline
$h_3=h_{125}$ &$t,\tau$ &$\tau$ &$t$ &  $\checkmark$&  $\checkmark$\\\hline
$h_4=h_{125}$ &$t,\tau$ &$\tau$ &$t$ &  $\checkmark$ &  $\checkmark$\\\hline
$h_5=h_{125}$ &$t,\tau$ &$\underline{\times}$ &$t$ & $\underline{\times}$
& $\underline{\times}$\\\hline
\end{tabular}
\caption{Current results for the large CP-odd $h_{125}bb$ couplings.
A check-mark means that, using all experiments,
$|c^o_{bb}| \gtrsim |c^e_{bb}|$ is achievable.
The underlined cross corresponds to cases where not a single valid point
was obtained. The entries with $\tau$  and/or $t$ are model choices that
have the ratio $c^o_{bb}/c^e_{bb}$ limited by the direct experimental bound
on $c^o_{\tau\tau}$ \cite{CMS:2021sdq,ATLAS:2022akr} and/or $c^o_{tt}$
\cite{ATLAS:2020ior}, respectively.
This Table should be compared with Table~3 in Ref.~\cite{Biekotter:2024ykp},
where the complex two-Higgs doublet model was addressed.
}
\label{tab:summary}
\end{table}

\subsection{\texorpdfstring{The $h_{125}=h_5$ ordering}{The h125=h5 ordering}}

The $h_{125}=h_5$ ordering was the most computationally difficult to get valid
points in,
and ultimately unrealized in Types II, Y, and Z.
The difficulty can be interpreted as a consequence of a more
restricted setup, as we required 4 neutral scalars lighter than the
observed $h_{125}$ to not be already observed with current experimental results. 
Nevertheless, we obtain viable parameter regions in Type I and X models
with the $h_{125}=h_5$ ordering, showing that this interesting scenario
is feasible for those Types.

\begin{figure}[htb]
  \centering
    \includegraphics[width=0.45\textwidth]{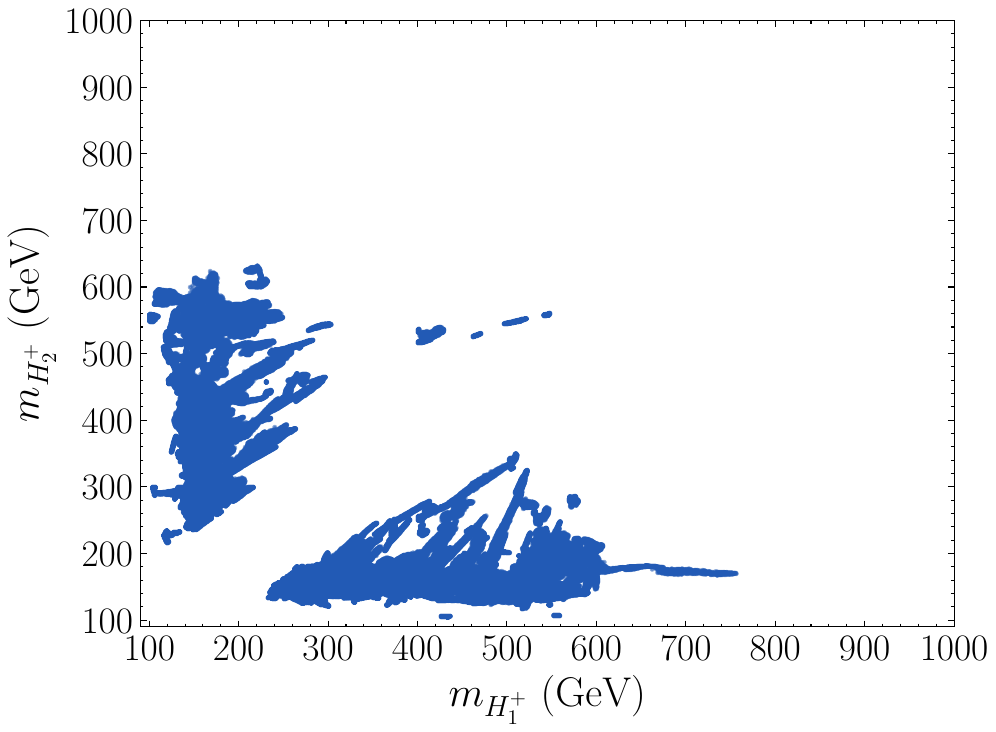}
  \caption{Allowed regions in the charged scalar masses plane for the
Type II model with the ordering $h_{125}=h_3$, including focused runs on the low charged scalar mass region.
}
  \label{fig:mC_t2_h3}
\end{figure}

To address the failure in Type II,
we start by looking at the viable charged scalar masses.
In the plot of Fig.~\ref{fig:mC_t2_h3}, the charged scalar masses plane is
shown for the Type II model with $h_{125}=h_3$ ordering.
From it, one can see that the region of simultaneously
low masses, $\lesssim 200\textrm{GeV}$, is excluded.
This feature,
is due to the $b \rightarrow s \gamma$ \cite{Akeroyd:2020nfj} bound.
This was also the situation found in Ref.~\cite{Boto:2021qgu},
for the real Type Z $\mathbb{Z}_3$-symmetric 3HDM 
-- see Figure~13 of Ref.~\cite{Boto:2021qgu}.
We have found this to occur also for Types Y and Z.

The complete failure of the $h_5 = h_{125}$ ordering in Types II, Y, and Z models can now be
explained by the clashing of three incompatible requests:
1) the $\Gamma(B\longrightarrow X_s\gamma)$
constraint requires at least one charged scalar mass greater
than the Higgs mass;
2) From the STU constraints,
it is necessary to have a neutral scalar with a mass similar to each charged scalar
\cite{Das:2022gbm};
3) all neutral scalars must be lighter than
$125\,\textrm{GeV}$.
This struggle could be seen in the evolution of
the ML algorithm, as all runs that could comply with all other constraints,
only had one of the two, $\Gamma(B\longrightarrow X_s\gamma)$ or STU,
also respected.
This demonstrates one powerful feature of monitored optimization algorithms in identifying
tensions between constraints imposed.

The different behavior in comparison to the Type I and X (Lepton Specific)
models are a result of the coupling combinations that enter the
$\Gamma(B\longrightarrow X_s\gamma)$ calculation. As described in
detail in Ref.~\cite{Akeroyd:2020nfj}, the relevant structures in
these two models do not differ much from the SM. 

Next we turn to the search for large CP-odd $h_{125}$ couplings to fermions
in the various Types and mass orderings.

\subsection{Type I}

In Type I models, the same doublet couples to all the fermions,
leading to equal $h_{125}$-fermion couplings 
(except for the minus sign in the up-type CP-odd coupling).
Thus, the constraints in the couplings on the fermions of a
specific electric charge directly restrain
the couplings of the other two possibilities.
In particular,
the possibility of a scenario with $|c^o| > |c^e|$ is excluded for
all Higgs fermion couplings based on the constraints on the top
quark coupling, arising from $tth$ production data~\cite{ATLAS:2020ior}.
It was possible to find viable parameter regions across all five
orderings with Type I Yukawa couplings.
The plots of Fig.~\ref{fig:ee_t1_h1,2,3,4,5} show
the obtained region for each ordering,
together with the experimental top constraints
from Ref.~\cite{ATLAS:2020ior}.
\begin{figure}[H]
\begin{subfigure}[b]{0.32\linewidth}
\centering
    \includegraphics[width=0.95\textwidth]{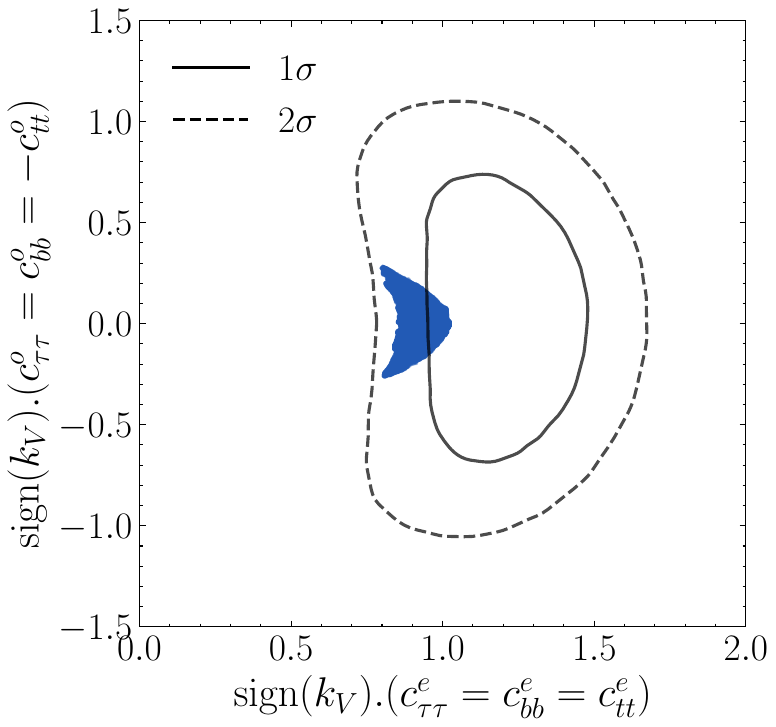}
\vspace*{3mm}
\caption{$h_{125}=h_1$.}
\end{subfigure}
\begin{subfigure}[b]{0.32\linewidth}
\centering
    \includegraphics[width=0.95\textwidth]{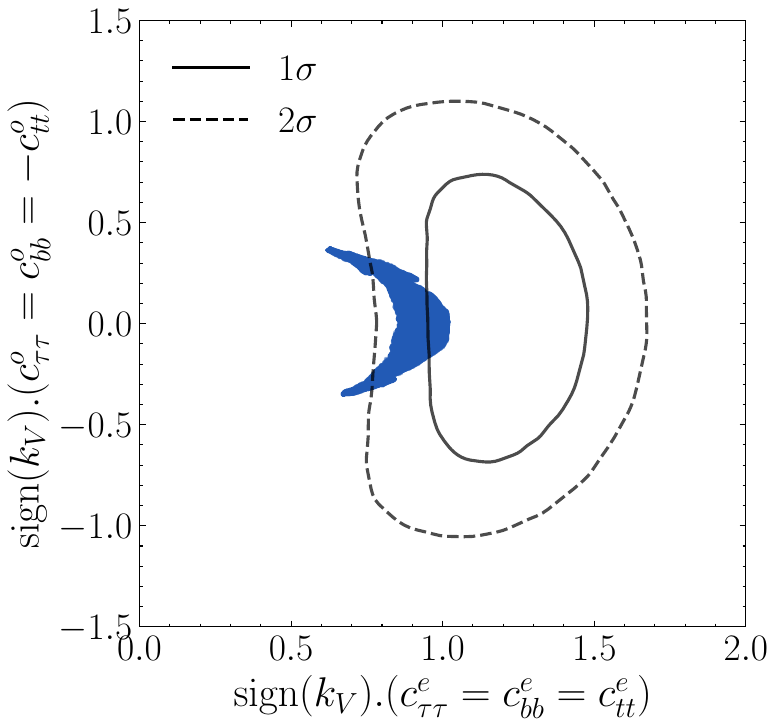}
\vspace*{3mm}
\caption{$h_{125}=h_2$.}
\end{subfigure}
\begin{subfigure}[b]{0.32\linewidth}
\centering
    \includegraphics[width=0.95\textwidth]{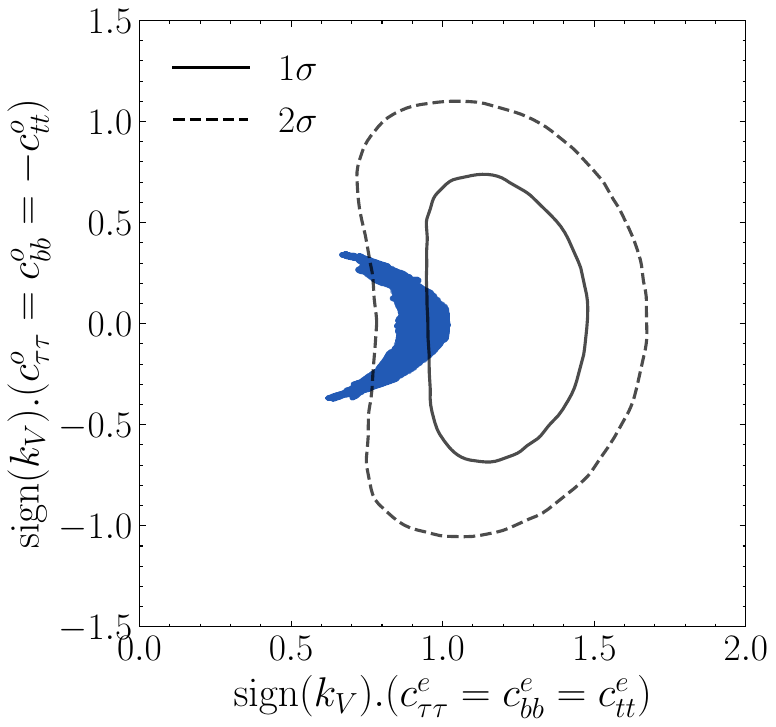}
\vspace*{3mm}
\caption{$h_{125}=h_3$.}
\end{subfigure}
\bigskip
\hspace{2.5cm}
\begin{subfigure}[b]{0.32\linewidth}
\centering
    \includegraphics[width=0.95\textwidth]{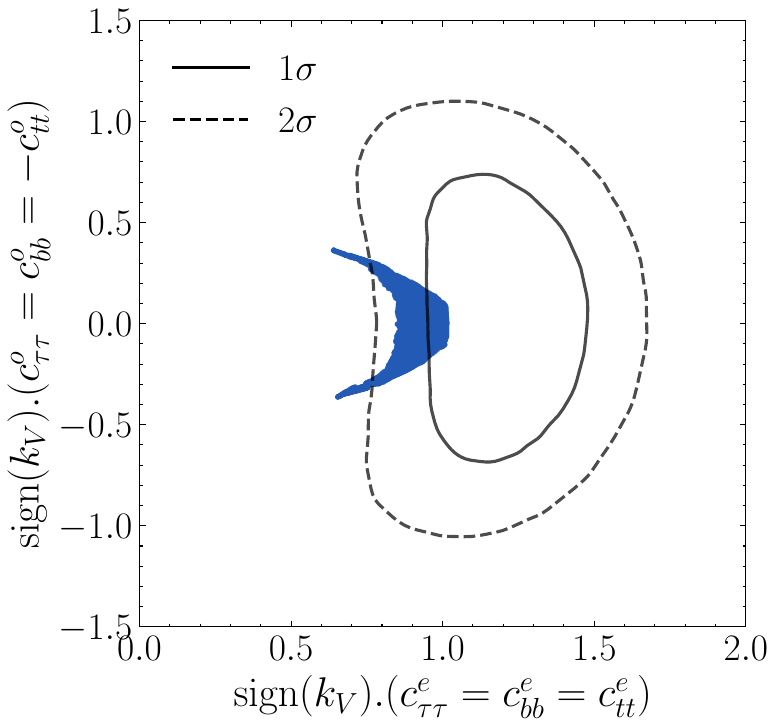}
\vspace*{3mm}
\caption{$h_{125}=h_4$.}
\end{subfigure}
\hfill
\begin{subfigure}[b]{0.32\linewidth}
\centering
    \includegraphics[width=0.95\textwidth]{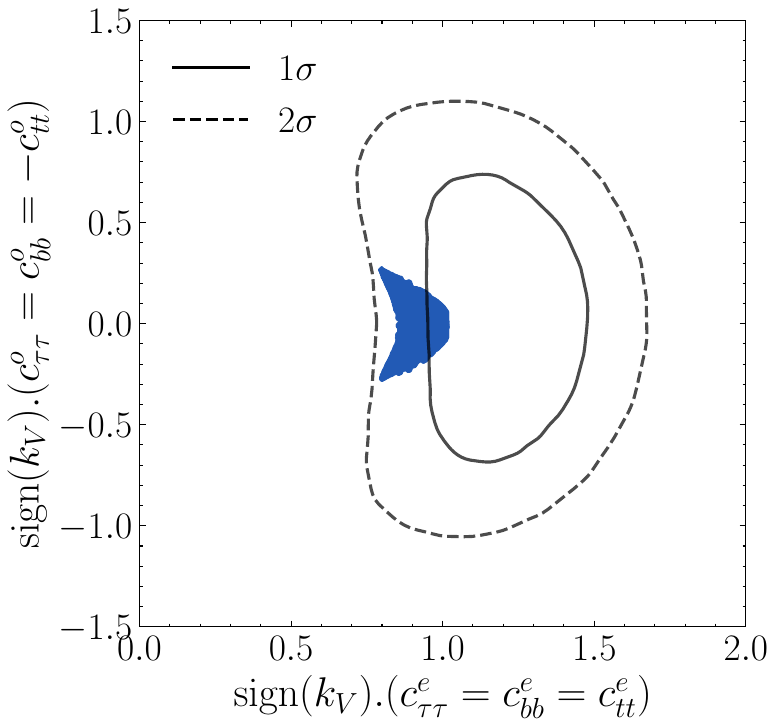}
\vspace*{3mm}
\caption{$h_{125}=h_5$.}
\end{subfigure}
\hspace{2.5cm}
\vspace*{-5mm}
  \caption{Allowed $h_{125}$-fermion coupling regions obtained in the Type I C3HDM
for the possible orderings.
The solid and dashed lines correspond to the limits on top-quark couplings from
Ref.~\cite{ATLAS:2020ior}, including
runs focused on the couplings shown.
}
  \label{fig:ee_t1_h1,2,3,4,5}
\end{figure}
In particular,
the Type I 3HDM allows, for all mass orderings,
the possibility of having $c^o_{tt}$ reach the $2\sigma$ line currently allowed by experiment.
This coincides with what was found previously for the case $h_{1}=h_{125}$ 
in the Type Z C3HDM \cite{deSouza:2025bpl}.
Notice from the figures that the wrong-sign solution is already excluded in Type I.

\subsection{Types II and X}
For the Types II and X, the Higgs lepton couplings are distinct from the other
fermion couplings, and thus, not directly constrained by the up-type quark
coupling constraints.
\begin{figure}[H]\centering
\begin{subfigure}[b]{0.34\linewidth}
\centering
    \includegraphics[width=0.9\textwidth]{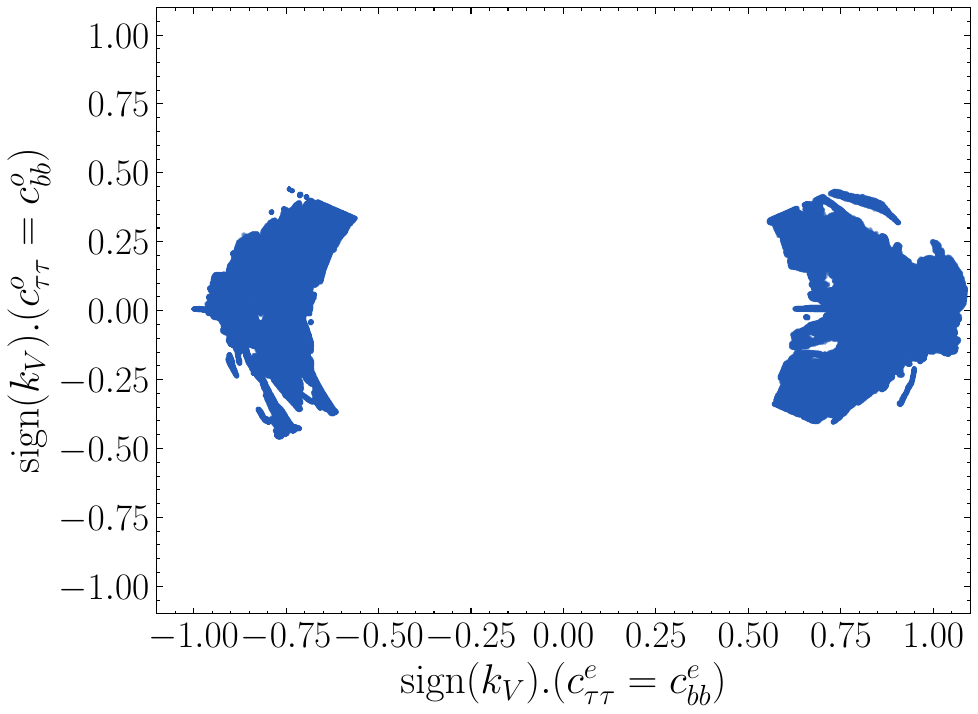}
\vspace*{3mm}
\caption{$h_{125}=h_1$.}
\end{subfigure}
\hspace{2cm}
\begin{subfigure}[b]{0.34\linewidth}
\centering
    \includegraphics[width=0.9\textwidth]{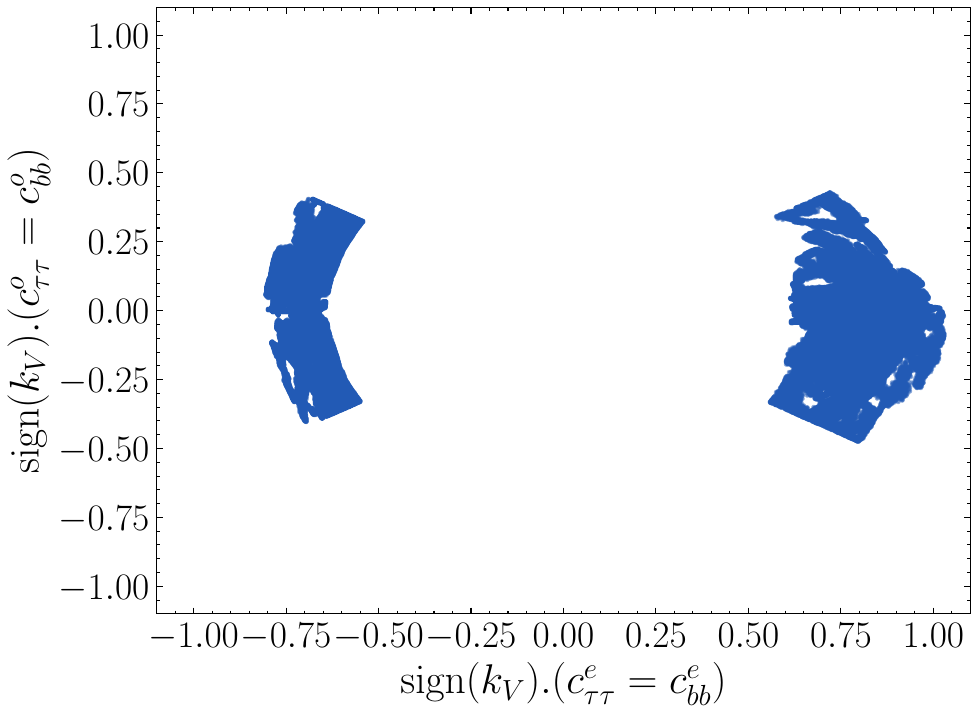}
\vspace*{3mm}
\caption{$h_{125}=h_2$.}
\end{subfigure}
\bigskip
\begin{subfigure}[b]{0.34\linewidth}
\centering
    \includegraphics[width=0.9\textwidth]{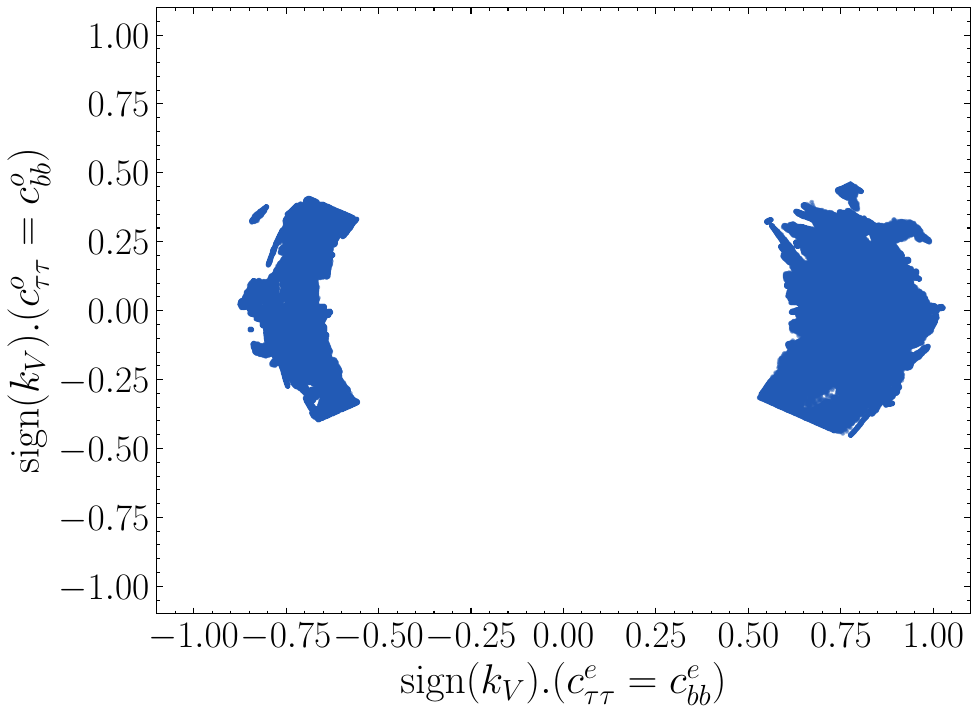}
\vspace*{3mm}
\caption{$h_{125}=h_3$.}
\end{subfigure}
\hspace{2cm}
\begin{subfigure}[b]{0.34\linewidth}
\centering
    \includegraphics[width=0.9\textwidth]{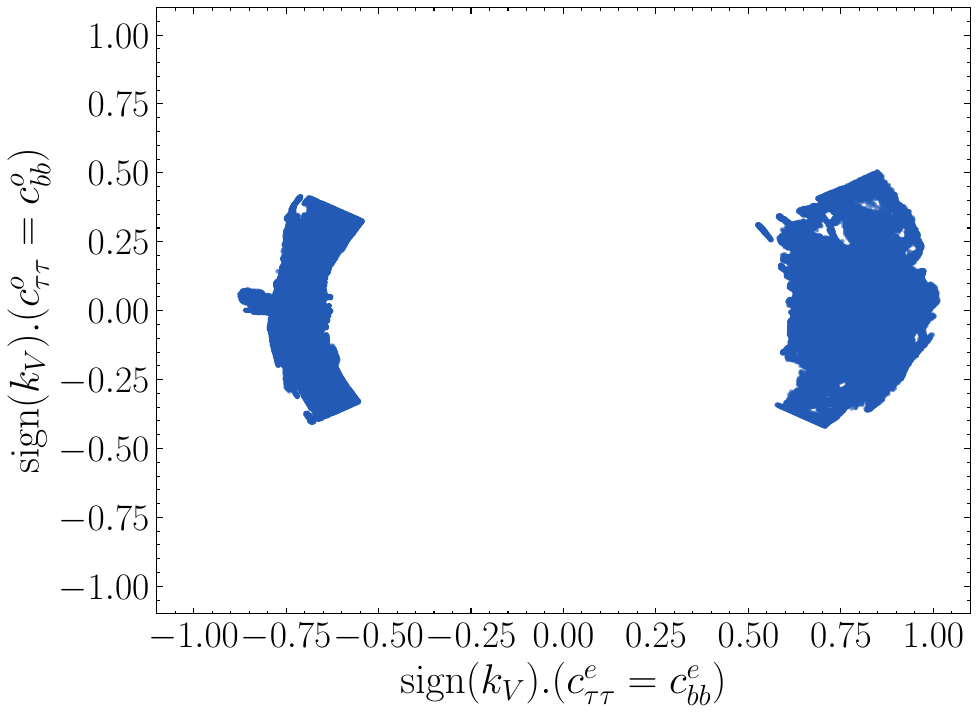}
\vspace*{3mm}
\caption{$h_{125}=h_4$.}
\end{subfigure}
  \caption{Allowed $h_{125}$-fermion coupling regions in Type II models for the different ordering choices, including runs focused on the couplings shown.
}
  \label{fig:ee_t2_h1,2,3,4}
\end{figure}
For Type II models, 
we do not find a single viable point for the $h_{125}=h_5$ ordering.
From the remaining cases, we obtain the possibility  $|c^o_\tau|\approx |c^e_\tau|$,
having points limited by the experimental bound on
$|\theta_{\tau}| = |\arctan(c^o_{\tau\tau} / c^e_{\tau\tau})| < 34^\circ$ 
\cite{CMS:2021sdq,ATLAS:2022akr}.
This is depicted in Fig.~\ref{fig:ee_t2_h1,2,3,4},
showing the viable parameter regions in the $c^e_{\tau\tau}-c^o_{\tau\tau}$ plane.
Notice that,
for the four surviving mass orderings of Type II,
there are many solutions around the wrong-sign point
$\textrm{sign}(k_V) c^e_{bb} \equiv \textrm{sign}(k_V) c^e_{\tau\tau} = -1$.

We now turn to the Type X models,
also known as Lepton Specific.
Fig.~\ref{fig:X_ee_t3_h1,2,3,4,5} depicts the
results for the $\tau\bar{\tau}$ couplings.
It was possible to find viable parameter regions for all five possible 
orderings, and this was computationally faster than for the Type II models.
The possibility of $|c^o_{\tau\tau}|\approx |c^e_{\tau\tau}|$
remains realizable and limited by the experimental
limit on $|\theta_{\tau}| = |\arctan(c^o_{\tau\tau} / c^e_{\tau\tau})| < 34^\circ$ 
\cite{CMS:2021sdq,ATLAS:2022akr}.
However,
the possibility of $|c^o_{bb}|\approx |c^e_{bb}|$ is excluded,
as the down-type and up-type quark couplings have the same absolute value.
Finally,
in Type X for the charged leptons (as in Type II),
there are many solutions around the wrong-sign point
$\textrm{sign}(k_V) c^e_{\tau\tau} = -1$.
\begin{figure}[H]
\begin{subfigure}[b]{0.32\linewidth}
\centering
    \includegraphics[width=0.95\textwidth]{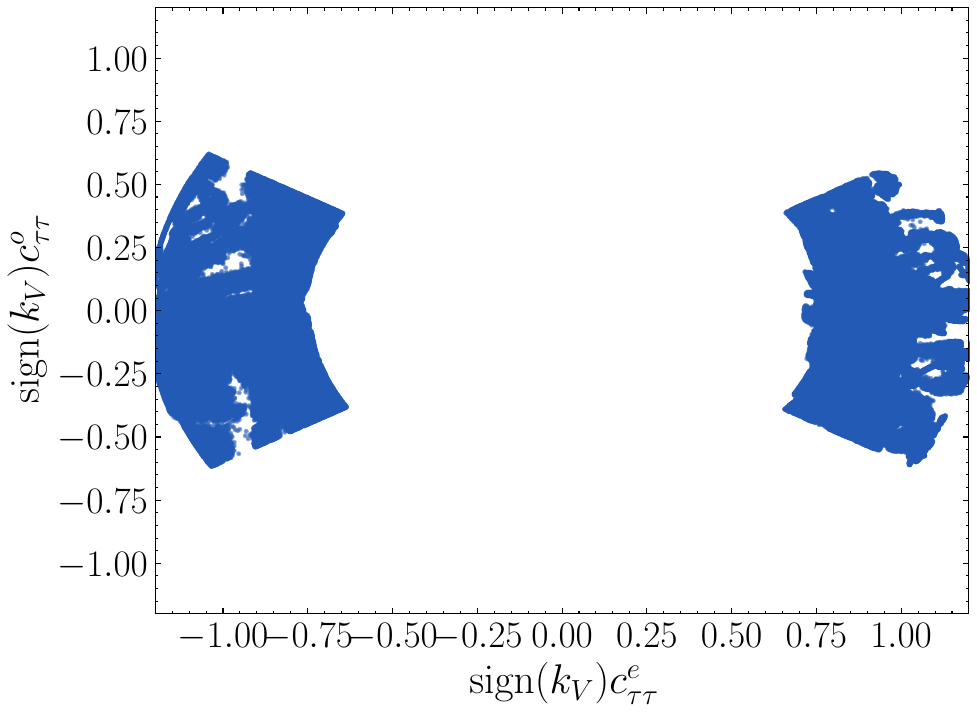}
\vspace*{3mm}
\caption{$h_{125}=h_1$.}
\end{subfigure}
\begin{subfigure}[b]{0.32\linewidth}
\centering
    \includegraphics[width=0.95\textwidth]{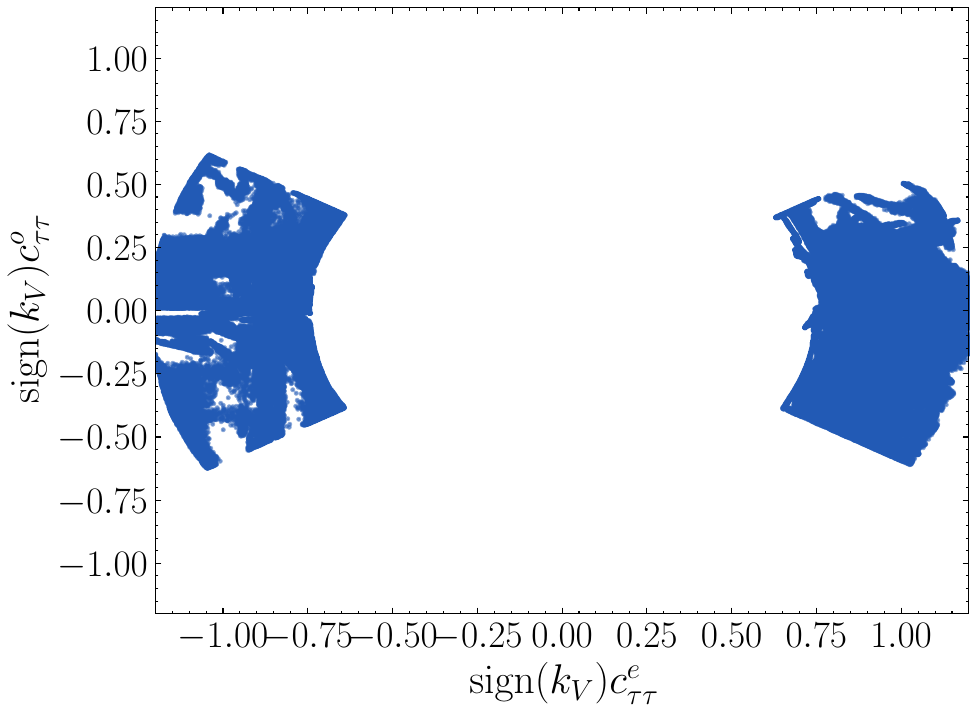}
\vspace*{3mm}
\caption{$h_{125}=h_2$.}
\end{subfigure}
\begin{subfigure}[b]{0.32\linewidth}
\centering
    \includegraphics[width=0.95\textwidth]{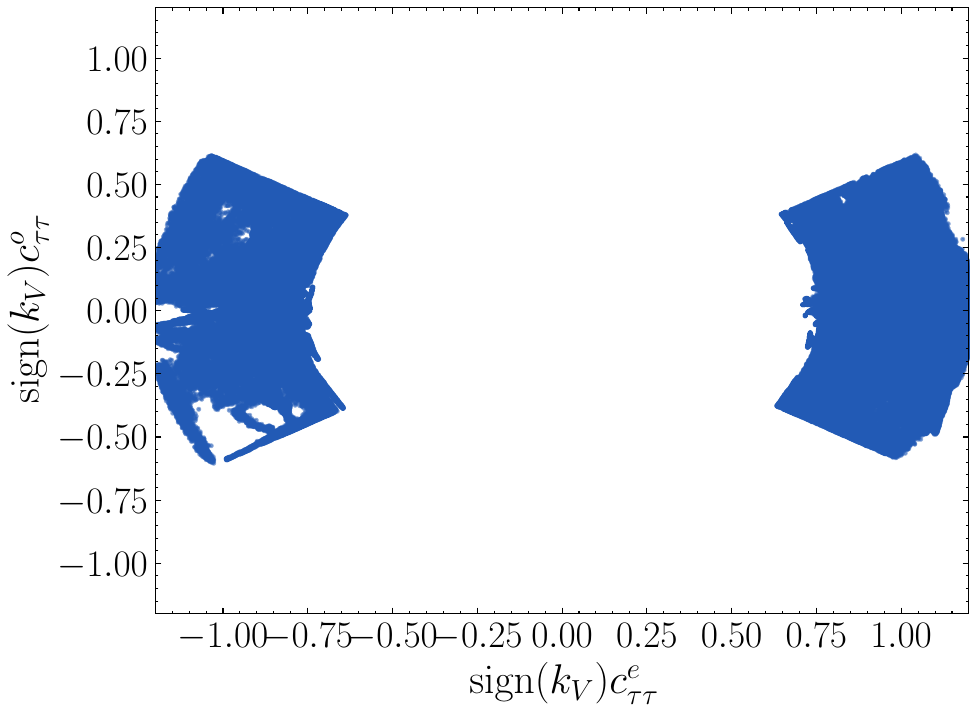}
\vspace*{3mm}
\caption{$h_{125}=h_3$.}
\end{subfigure}
\bigskip
\hspace{2.5cm}
\begin{subfigure}[b]{0.32\linewidth}
\centering
    \includegraphics[width=0.95\textwidth]{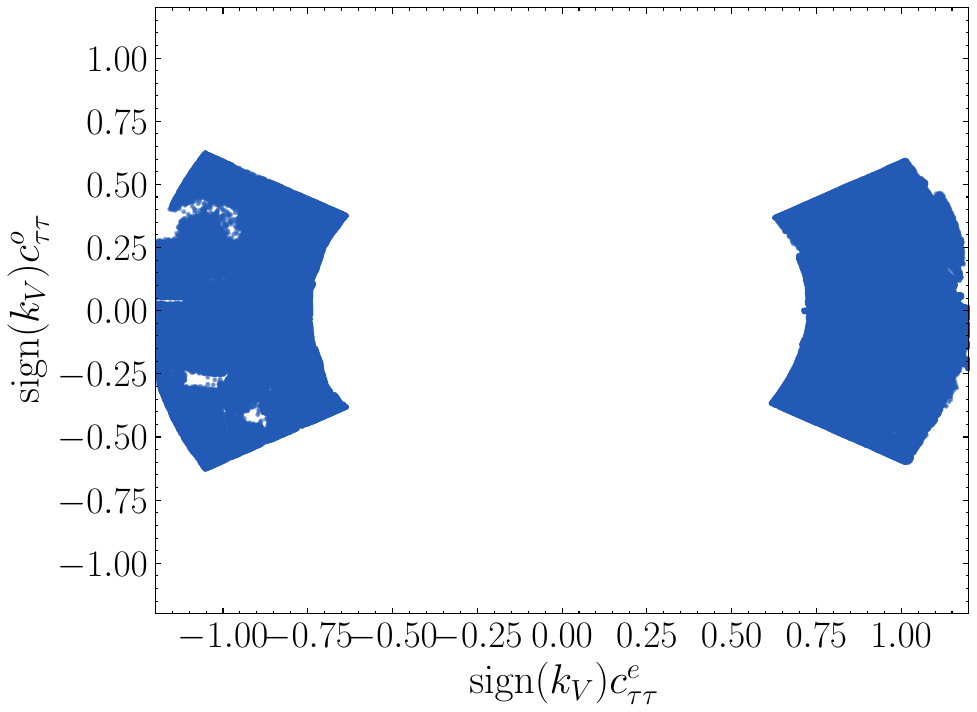}
\vspace*{3mm}
\caption{$h_{125}=h_4$.}
\end{subfigure}
\hfill
\begin{subfigure}[b]{0.32\linewidth}
\centering
    \includegraphics[width=0.95\textwidth]{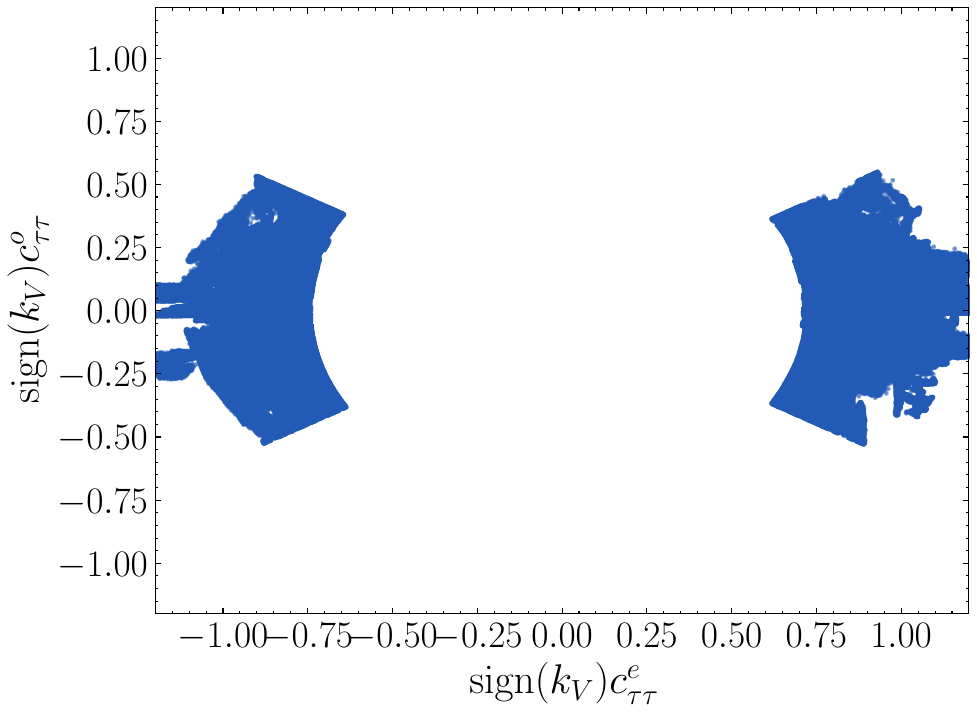}
\vspace*{3mm}
\caption{$h_{125}=h_5$.}
\end{subfigure}
\hspace{2.5cm}
\vspace*{-5mm}
  \caption{Allowed $h_{125}$-fermion coupling regions in Type X models for the different ordering choices, including runs focused on the couplings shown.
}
  \label{fig:X_ee_t3_h1,2,3,4,5}
\end{figure}

\subsection{Types Y and Z}

For the Type Y and Z models, the down-type quark couplings are not directly
limited by experimental constraints on the top and/or tau,
opening the possibility of having a purely pseudoscalar coupling,
with $c^e_{bb}=0$.
In both Types,
it was possible to acquire viable point regions for all orderings except $h_{125}=h_5$.
The Type Z model with $h_{125}=h_1$ ordering was the sole focus
of Refs.~\cite{Boto:2024jgj,deSouza:2025bpl},
and thus not repeated in this work.
For both models, the wrong-sign region is also completely populated.
Indeed, albeit with some gaps to fill with longer simulations,
the ``ellipses'' in the $\textrm{sign}(k_V)c^e_{bb}$- $\textrm{sign}(k_V)c^o_{bb}$
plane are complete.
\begin{figure}[H]\centering
\begin{subfigure}[b]{0.34\linewidth}
\centering
    \includegraphics[width=0.9\textwidth]{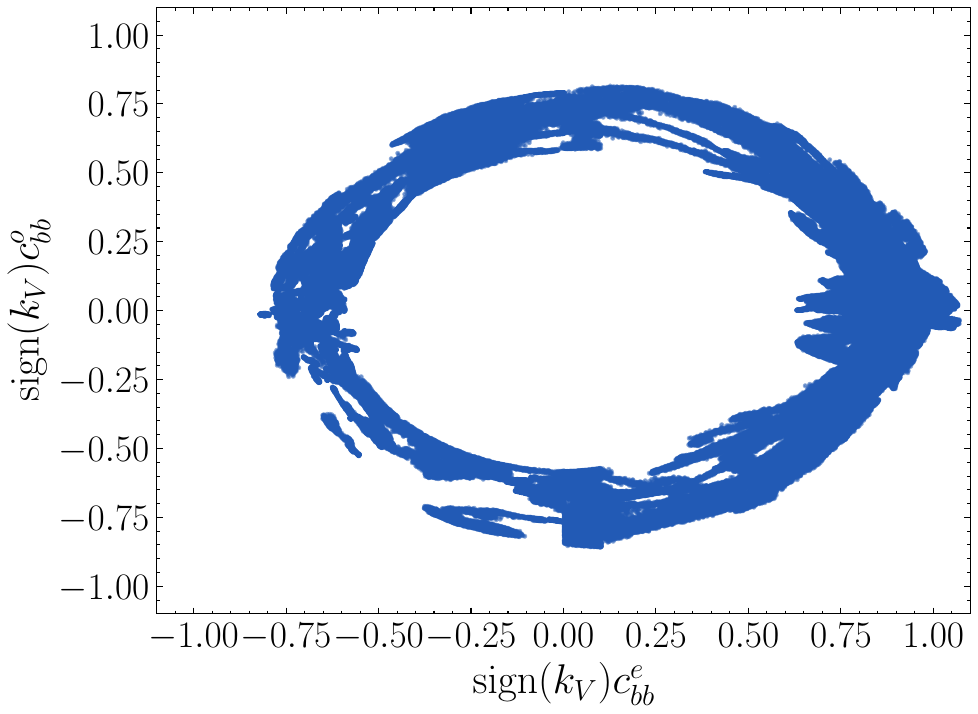}
\vspace*{3mm}
\caption{$h_{125}=h_1$.}
\end{subfigure}
\hspace{2cm}
\begin{subfigure}[b]{0.34\linewidth}
\centering
    \includegraphics[width=0.9\textwidth]{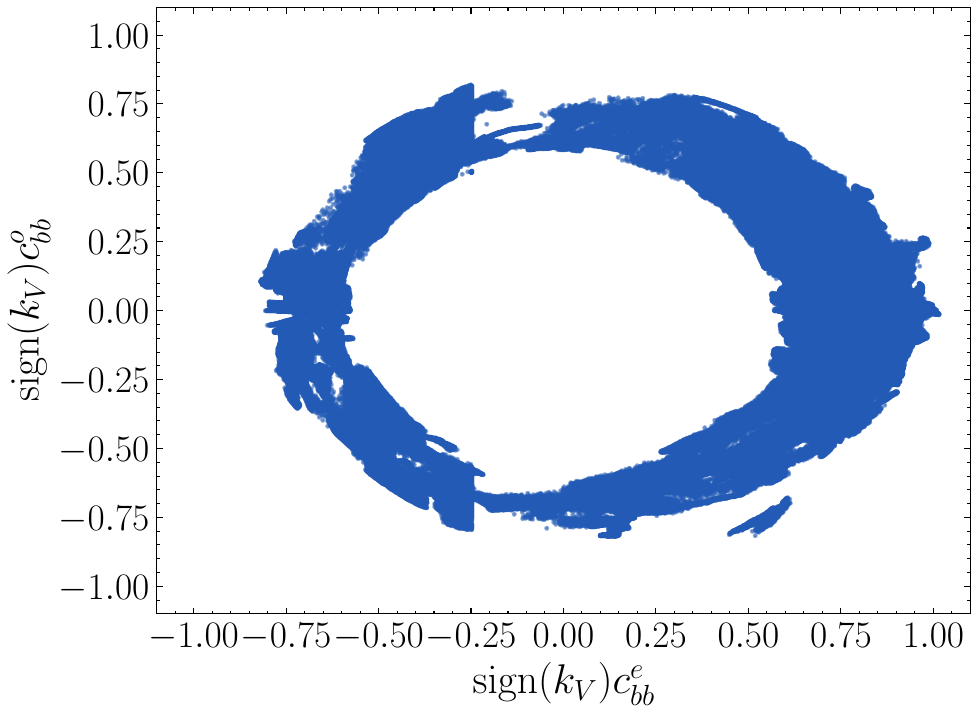}
\vspace*{3mm}
\caption{$h_{125}=h_2$.}
\end{subfigure}
\bigskip
\begin{subfigure}[b]{0.34\linewidth}
\centering
    \includegraphics[width=0.9\textwidth]{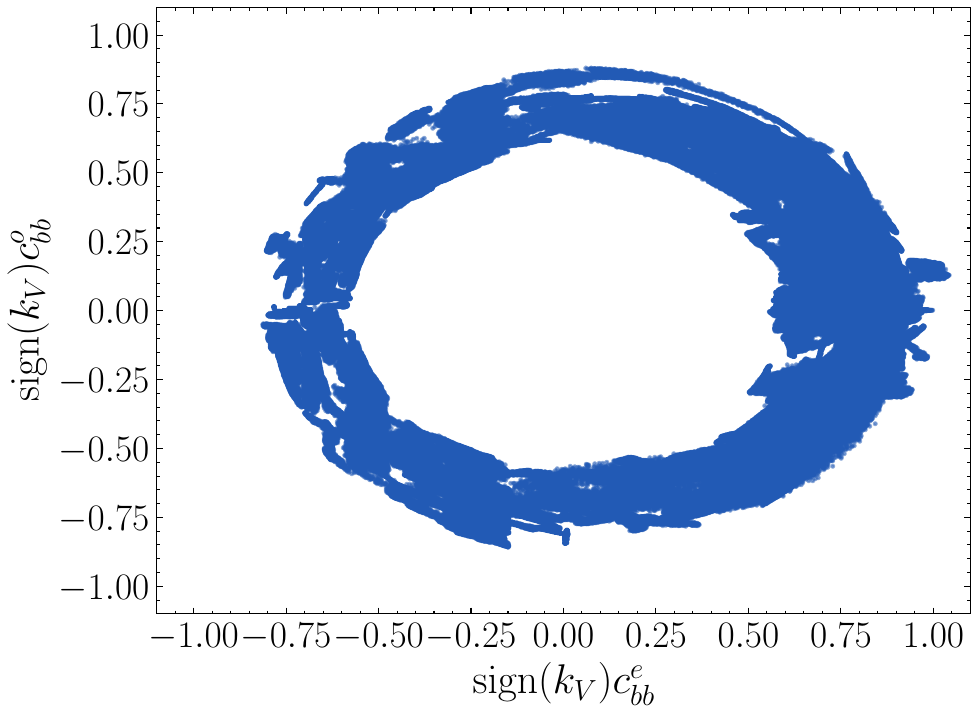}
\vspace*{3mm}
\caption{$h_{125}=h_3$.}
\end{subfigure}
\hspace{2cm}
\begin{subfigure}[b]{0.34\linewidth}
\centering
    \includegraphics[width=0.9\textwidth]{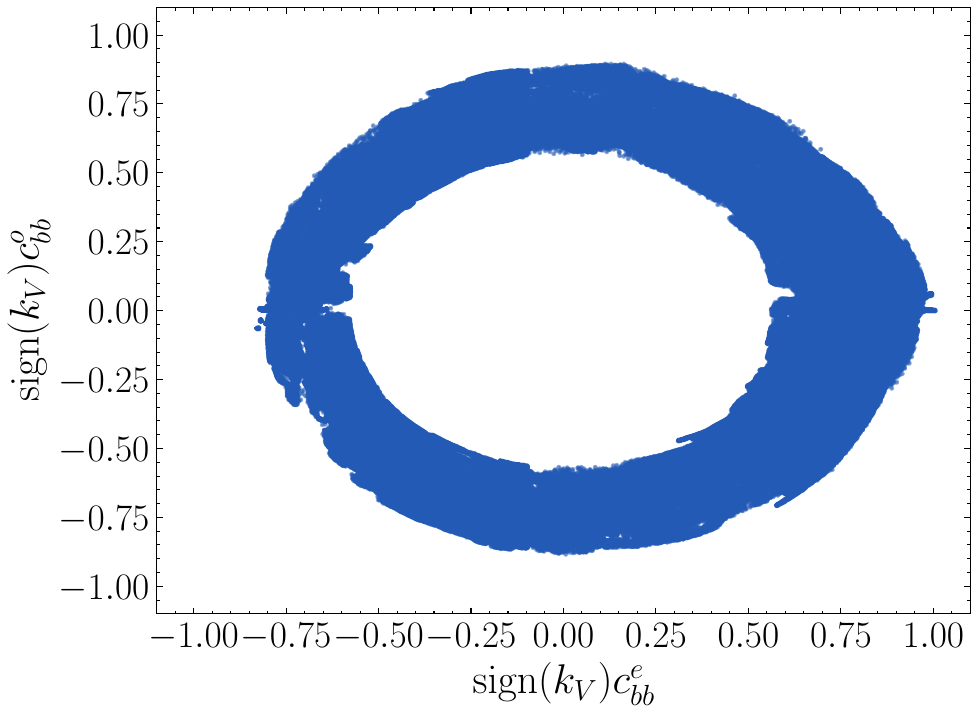}
\vspace*{3mm}
\caption{$h_{125}=h_4$.}
\end{subfigure}
  \caption{Allowed $h_{125}$-fermion coupling regions in Type Y models for the different ordering choices, including runs focused on the couplings shown.
}
  \label{fig:bb_t4_h1,2,3,4}
\end{figure}
The results for the $b\bar{b}$ coupling in Type Y models
are shown in Fig.~\ref{fig:bb_t4_h1,2,3,4},
highlighting the possibility of a pure CP-odd coupling
between the $h_{125}$ and the down-type quarks.
The allowed region shows the freedom gained from extending the C2HDM model,
in which this possibility was excluded in \cite{Biekotter:2024ykp}.
However, for models of this Type, the region with
$|c^o_{\tau\tau}|\approx |c^e_{\tau\tau}|$ is excluded,
since the lepton and up-type quark Higgs couplings are equal,
enforcing the up-type quark couplings' tight restrictions on the lepton couplings,
resembling the plots of Type I models.
\begin{figure}[H]
  \centering
  \begin{tabular}{ccc}
    \includegraphics[width=0.32\textwidth]{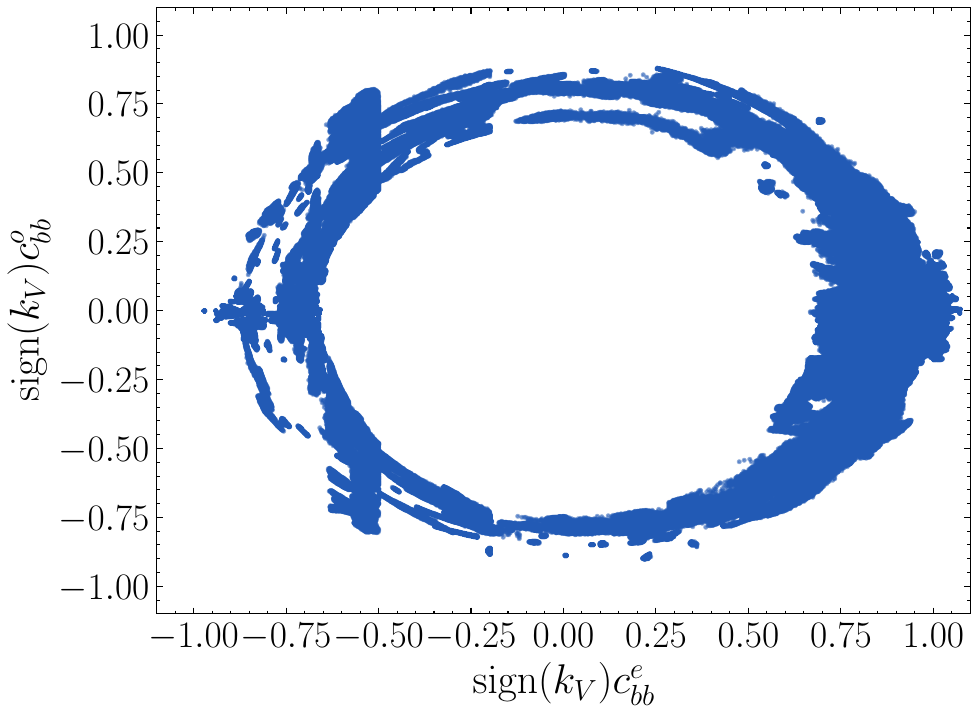}
    &
    \includegraphics[width=0.32\textwidth]{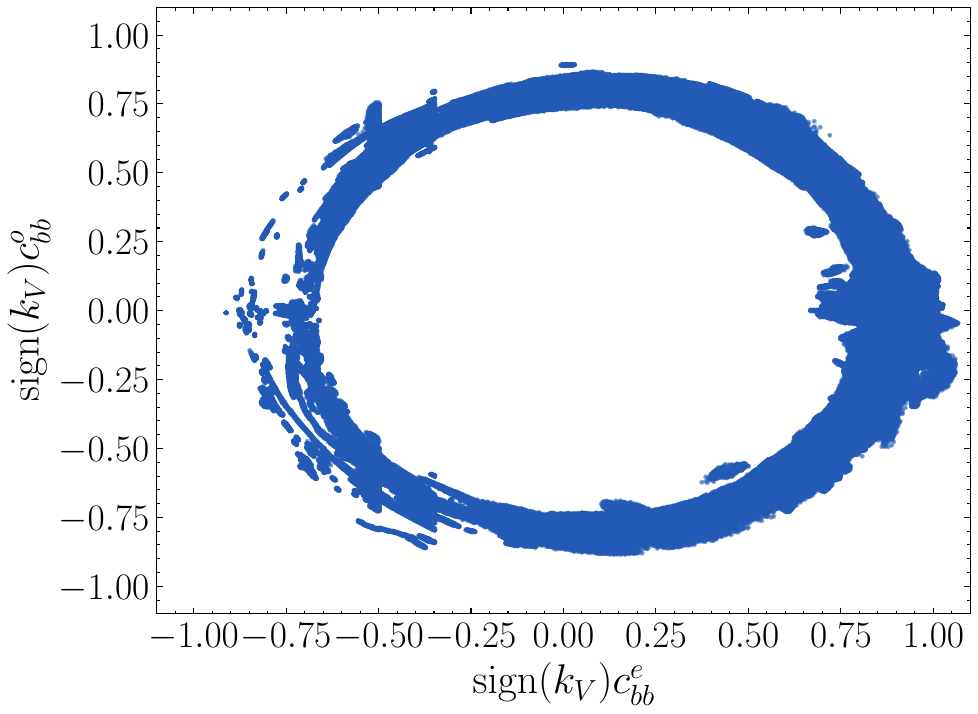}
    &
    \includegraphics[width=0.32\textwidth]{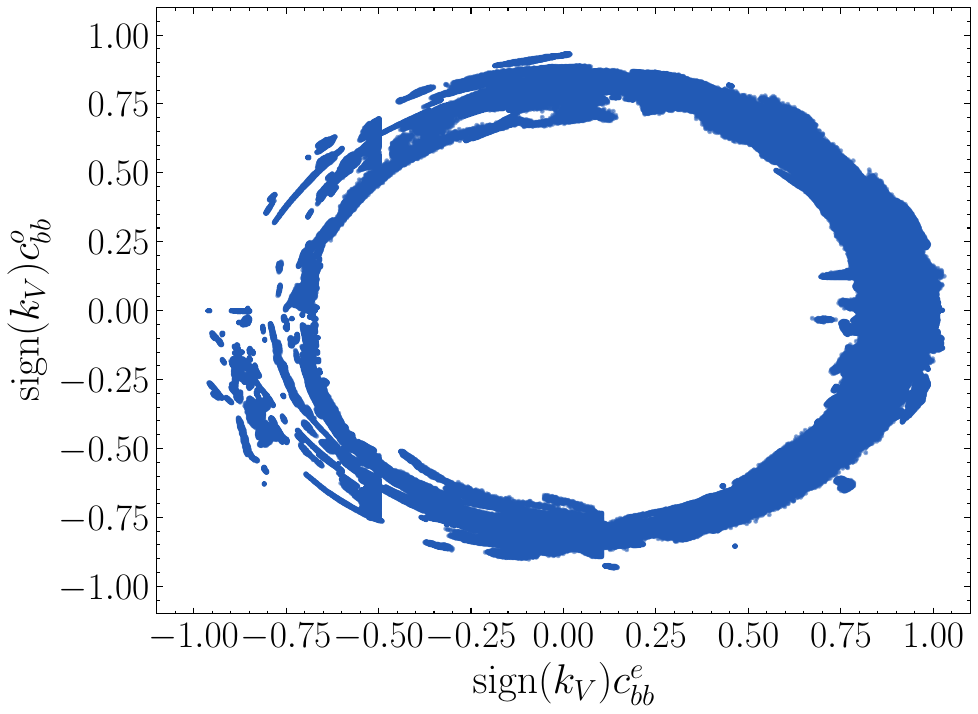}
  \end{tabular}
  \caption{Allowed $h_{125}$-down-type quark coupling regions
in Type Z models with the orderings $h_{125}=h_2$, $h_3$ and $h_4$, from left to right, respectively. Includes focused runs with novelty reward on the couplings $c^e_{bb}-c^o_{bb}$.
}
  \label{fig:bb_t5_h2,3,4}
\end{figure}

For the Type Z models, Fig.~\ref{fig:bb_t5_h2,3,4} shows the
possible values for the down-type quark couplings compatible with all
the constraints, leading to conclusions similar to the Type Y models
in this plane. As for the lepton couplings, in this model,
they are also independent from the other two types of fermion couplings,
and, thus, there is the possibility of having
$|c^o_{\tau\tau}|\approx |c^e_{\tau\tau}|$,
consistent with the direct experimental bound in Eq.~\eqref{theta_tau}.

Fig.~\ref{fig:ee_t5_h2,3,4} shows the valid regions
for $h_{125}$-lepton coupling regions in Type Z models
with the orderings $h_{125}=h_2$, $h_3$ and $h_4$, from left to right, respectively.
\begin{figure}[H]
  \centering
  \begin{tabular}{ccc}
    \includegraphics[width=0.32\textwidth]{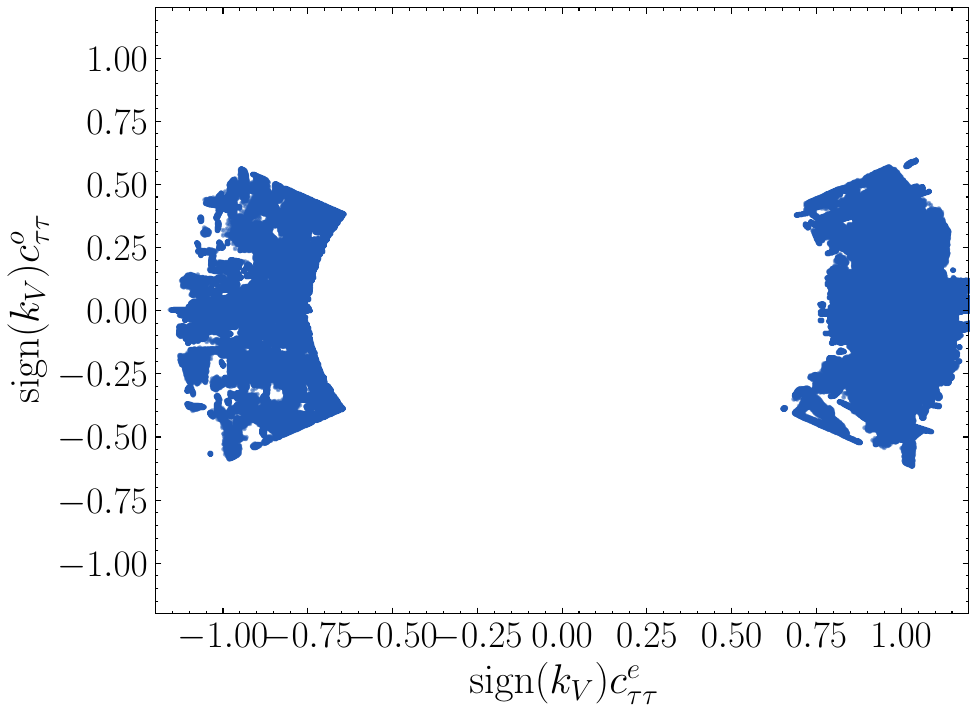}
    &
    \includegraphics[width=0.32\textwidth]{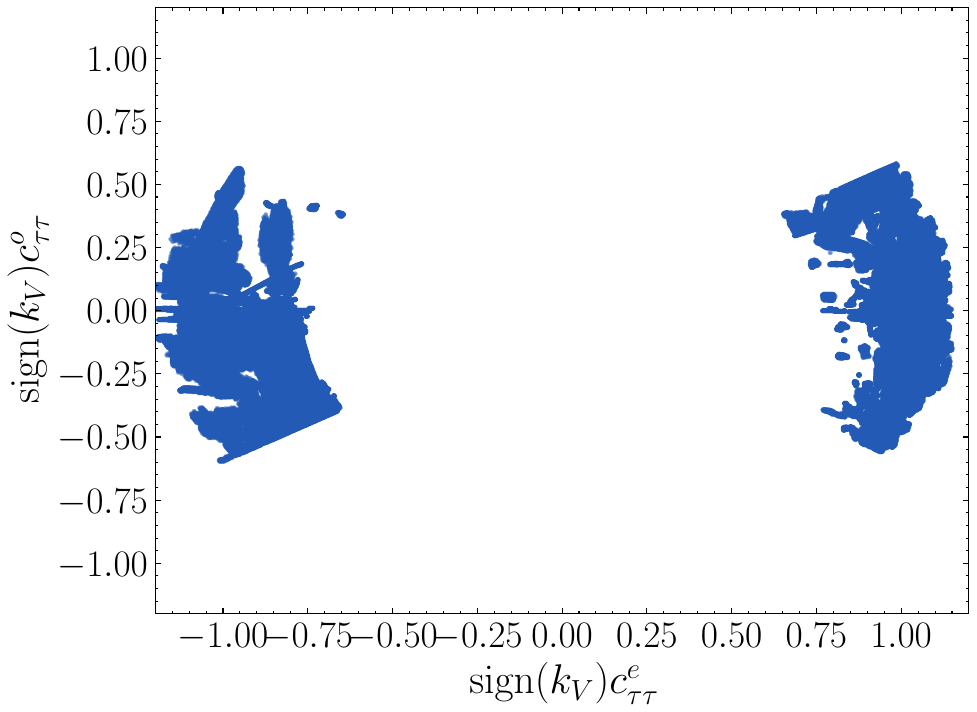}
    &
    \includegraphics[width=0.32\textwidth]{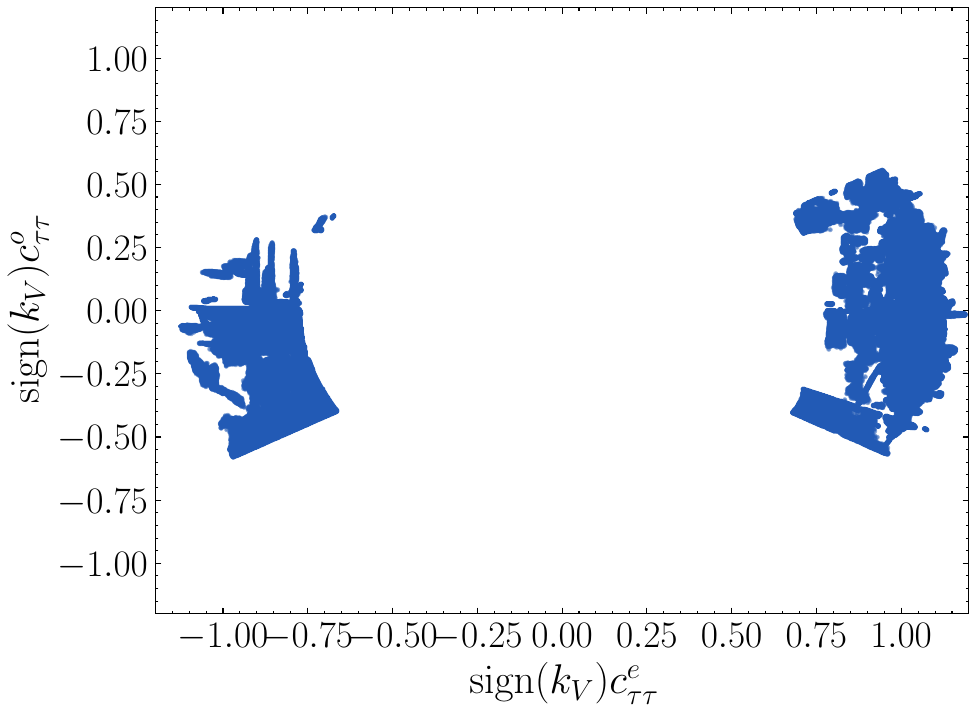}
  \end{tabular}
  \caption{Allowed $h_{125}$-lepton coupling regions in Type Z models
with the orderings $h_{125}=h_2$, $h_3$ and $h_4$, from left to right, respectively.  Includes focused runs with novelty reward on the couplings $c^e_{\tau\tau}-c^o_{\tau\tau}$.
%The colour code of the points is the same as in previous plots.
}
  \label{fig:ee_t5_h2,3,4}
\end{figure}

\subsection{\label{subsec:scalarornot}Is the 125GeV Higgs scalar or pseudoscalar?}

\begin{figure}[H]
  \centering
    \includegraphics[width=0.45\textwidth]{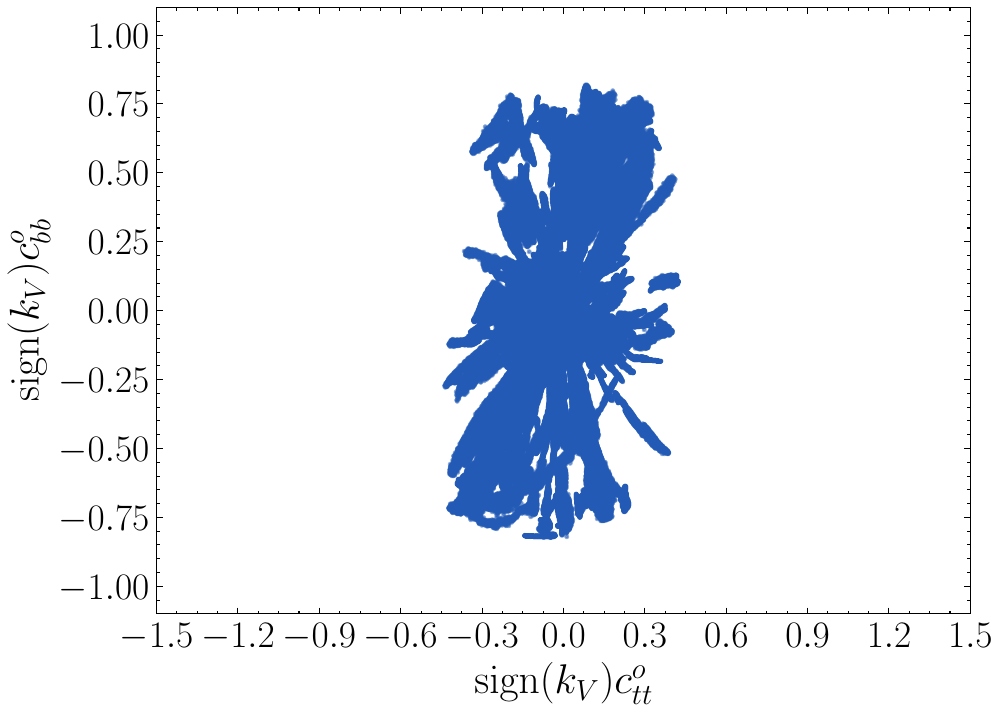}
  \caption{Allowed regions in the
$\textrm{sign}(k_V)c^o_{tt}$- $\textrm{sign}(k_V)c^o_{bb}$
plane for the
Type Y model with the ordering $h_{125}=h_2$, without runs with novelty reward on the plane shown.
}
  \label{fig:ttvsbb_t4_h2}
\end{figure}

Having found out that, for Types Y and Z,
and for all mass orderings except $h_{125}=h_5$,
one can have maximal CP-odd $h_{125}bb$ couplings,
we now turn to the question raised in Ref.~\cite{Fontes:2015xva}
of whether one can have $h_{125}$ couple as a pure scalar to some fermions,
while it couples as a pure pseudoscalar to others.
The situation is the same for all eight cases, so we concentrate on the
Type Y with $h_{125}=h_2$ ordering shown in Fig.~\ref{fig:ttvsbb_t4_h2}.
First, let us concentrate on the vertical line $\textrm{sign}(k_V)c^o_{tt}=0$,
in which $h_{125}$ couples to the top quark as a pure scalar.
We see that $\textrm{sign}(k_V)c^o_{bb}$ can be as large as $0.75$,
consistently with Fig.~\ref{fig:bb_t4_h1,2,3,4}(b),
meaning that $h_{125}$ couples to the bottom quark as a pure pseudoscalar.
It is extraordinary that,
after so much has been learned about the $125\textrm{GeV}$ Higgs particle,
such a peculiar situation is still possible.

Equally extraordinary is the complementary situation corresponding to the 
horizontal line $\textrm{sign}(k_V)c^o_{bb}=0$,
in which $h_{125}$ couples to the bottom quark as a pure scalar.
In this case,
the CP-odd $h_{125}$ coupling to the top quark is limited experimentally \cite{ATLAS:2020ior}
by Eq.~\eqref{theta_t}.
However, on the theoretical side,
this coupling can be as large as experimentally allowed.
Reversing the conclusion,
any small improvement on the bounds in Eq.~\eqref{theta_t} yields true knowledge
about this class of models.

In Type Z, we see the same features concerning the top versus bottom couplings as shown here for Type Y.
Besides, in Type Z, one can also have simultaneously large bottom and tau CP-odd couplings.
This is true for all orderings (except $h_{125}=h_5$), and can already be seen for
the special ordering $h_1=h_{125}$ on the left panel
of Figure~5 in Ref.~\cite{deSouza:2025bpl}.

\section{\label{sec:concl}Conclusions}

The experimental results in Eqs.~\eqref{theta_t} and \eqref{theta_tau}
still allow for a roughly equal CP-odd and CP-even component of the $125\textrm{GeV}$
Higgs' couplings to the top quark and to the tau lepton.
In contrast,
there is no direct experimental bound on the CP nature of the $h_{125}bb$
couplings.
This spurs the peculiar prospect that the Higgs found at LHC would have
a purely scalar $h_{125}tt$ coupling,
while it has a purely pseudoscalar $h_{125}bb$ coupling.
This possibility was first proposed in the context of the C2HDM \cite{Fontes:2015mea},
but recent experimental data precludes it \cite{Biekotter:2024ykp}.
This possibility was recovered in the C3HDM \cite{Boto:2024jgj,deSouza:2025bpl},
with Type Z couplings and taking $h_{125}$ to be the lightest of the
five scalars in such theories.

The question arises of whether this is a peculiar property of Type Z models and/or
the special mass ordering taken.
This issue is fully explored here, by a thorough analysis of
all twenty five possibilities arising out of the five Types and five orderings.
As shown in \cite{deSouza:2025bpl},
these models, having large parameter spaces, cannot be probed efficiently with
conventional scanning techniques.
An evolutionary strategy-based machine learning algorithm is employed
to significantly improve sampling efficiency,
with convergence toward valid regions accelerated by
a novelty-driven reward mechanism.
In order to fully explore the available predictions for the
CP nature of the bottom and tau $h_{125}$ couplings,
we performed our runs mostly with focus on the
$\textrm{sign}(k_V)c^e_{bb}$- $\textrm{sign}(k_V)c^o_{bb}$
or the
$\textrm{sign}(k_V)c^e_{\tau\tau}$- $\textrm{sign}(k_V)c^o_{\tau\tau}$
planes.
Our results are summarized in Table~\ref{tab:summary}.

We show that (regardless of the exact nature of the fermionic couplings) no single point 
can be found for the $h_5=h_{125}$ ordering in Types II, Y, and Z,
due to a confluence of competing experimental constraints, which we explain in detail.
Moreover, given the experimental bounds in
Eqs.~\eqref{theta_t} and \eqref{theta_tau},
there can be no pure pseudoscalar couplings in models
where the bottom couplings mirror those of the top
and/or tau.
This leaves eight possibilities: the lowest four orderings of Types Y and Z.
Remarkably, in all these cases, one can indeed have purely pseudoscalar
$h_{125}bb$ coupling, while having a purely scalar $h_{125}tt$ coupling.
We also found another unusual possibility: that the $h_{125}bb$ coupling is purely scalar,
while the CP-odd $h_{125}tt$ coupling is as large as allowed by current experiments.
We hope that our work helps to build the case for more precise experimental probes
into the CP nature of the $125\textrm{GeV}$ Higgs.

\section*{Acknowledgments}
\noindent
This work is supported in part by the Portuguese
Fundação para a Ciência e Tecnologia (FCT) through the PRR (Recovery and Resilience
Plan), within the scope of the investment "RE-C06-i06 - Science Plus
Capacity Building", measure "RE-C06-i06.m02 - Reinforcement of
financing for International Partnerships in Science,
Technology and Innovation of the PRR", under the project with
reference 2024.01362.CERN.
The work of the authors is
also supported by FCT under Contracts UID/00777/2025 (https://doi.org/10.54499/UID/00777/2025).
The FCT projects are partially funded through
POCTI (FEDER), COMPETE, QREN, and the EU.
The work of R. Boto is also supported by FCT with the PhD
grant PRT/BD/152268/2021.

\appendix

\section{\label{app:top}Top quark couplings}
This appendix is dedicated to the results for
the various Types and orderings with valid parameter space points,
shown on the $\textrm{sign}(k_V)c^e_{tt}$- $\textrm{sign}(k_V)c^o_{tt}$ plane.
Recall that the constraints on the CP-odd component of the top coupling
in Eq.~\eqref{theta_t} were \textit{not} used as constraints,
as expressed in Sec.~\ref{sec:scan}.

Recall that in Type I all fermions have the same couplings, and these have been shown
in Fig.~\ref{fig:ee_t1_h1,2,3,4,5}
We show in 
Figs.~\ref{fig:tt_t2_h1,2,3,4,5},
Figs.~\ref{fig:ee_t3_h1,2,3,4,5},
Figs.~\ref{fig:tt_t4_h1,2,3,4,5},
and Figs.~\ref{fig:tt_t5_h2,3,4}
the results for Types II, X, Y, and Z, respectively.
The results for the ordering $h_{125}=h_1$ in Type Z was studied in
\cite{Boto:2024jgj,deSouza:2025bpl} and is not repeated here.

One can see that the current $2\sigma$ limits are already impinging on the allowed parameter
space in most cases.
This is a strong motivation for improving the measurements of the CP-odd components
of the $h_{125}tt$ couplings.
Part of the aim of this separate appendix is precisely to help spur such efforts.

\begin{figure}[H]\centering
\begin{subfigure}[b]{0.34\linewidth}
\centering
    \includegraphics[width=0.9\textwidth]{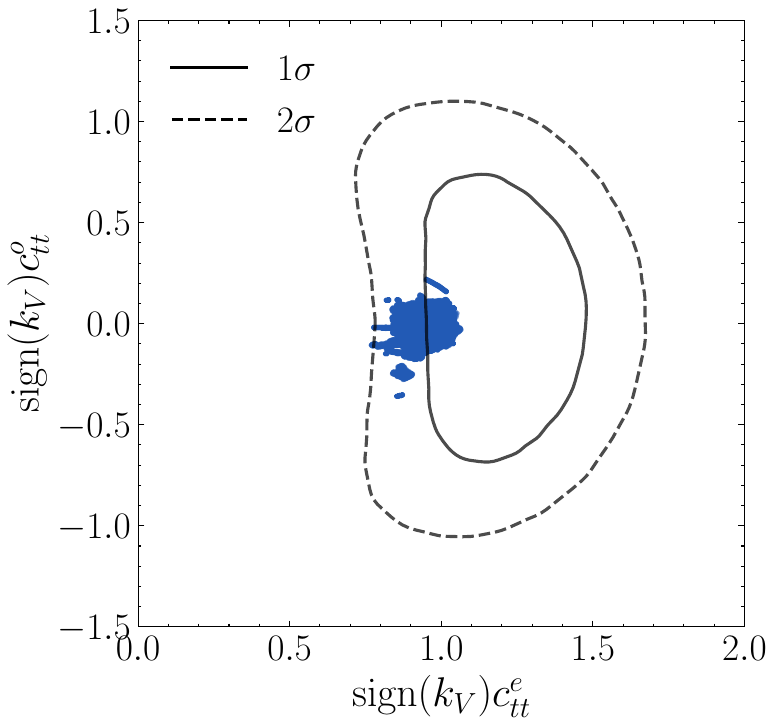}
\vspace*{3mm}
\caption{$h_{125}=h_1$.}
\end{subfigure}
\hspace{2cm}
\begin{subfigure}[b]{0.34\linewidth}
\centering
    \includegraphics[width=0.9\textwidth]{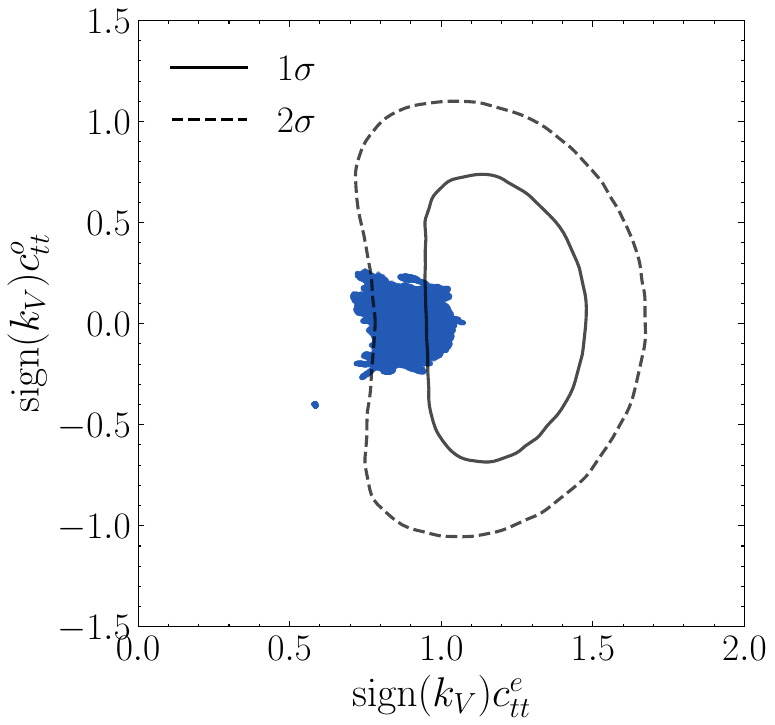}
\vspace*{3mm}
\caption{$h_{125}=h_2$.}
\end{subfigure}
\bigskip
\begin{subfigure}[b]{0.34\linewidth}
\centering
    \includegraphics[width=0.9\textwidth]{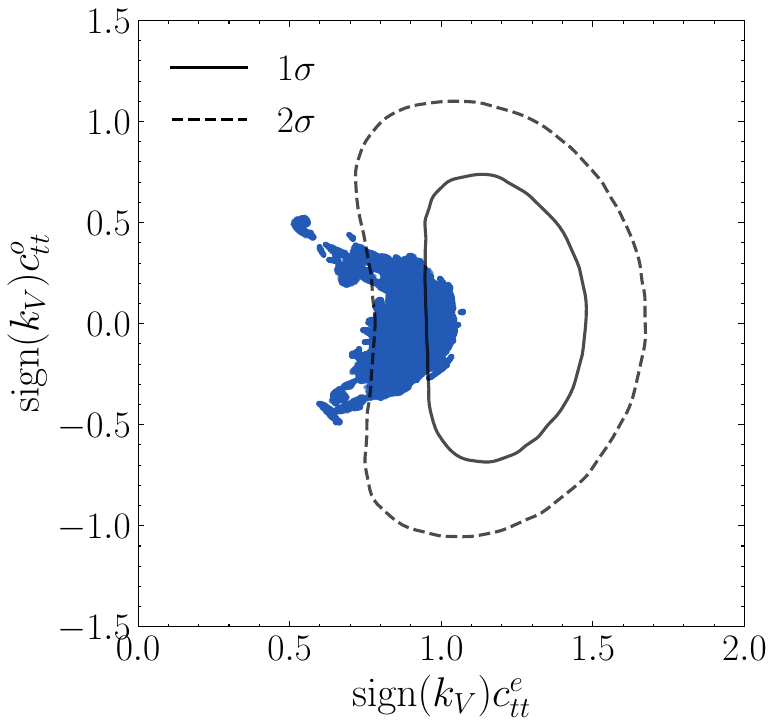}
\vspace*{3mm}
\caption{$h_{125}=h_3$.}
\end{subfigure}
\hspace{2cm}
\begin{subfigure}[b]{0.34\linewidth}
\centering
    \includegraphics[width=0.9\textwidth]{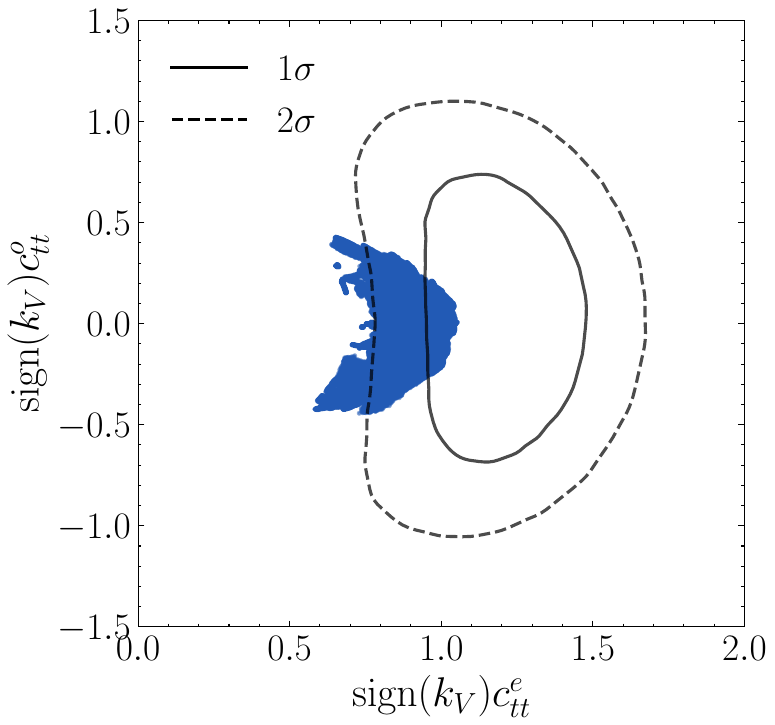}
\vspace*{3mm}
\caption{$h_{125}=h_4$.}
\end{subfigure}
  \caption{Allowed $h_{125}$-top coupling regions in Type II models, without novelty reward on the plane shown.
}
  \label{fig:tt_t2_h1,2,3,4,5}
\end{figure}

\begin{figure}[H]
\begin{subfigure}[b]{0.32\linewidth}
\centering
    \includegraphics[width=0.95\textwidth]{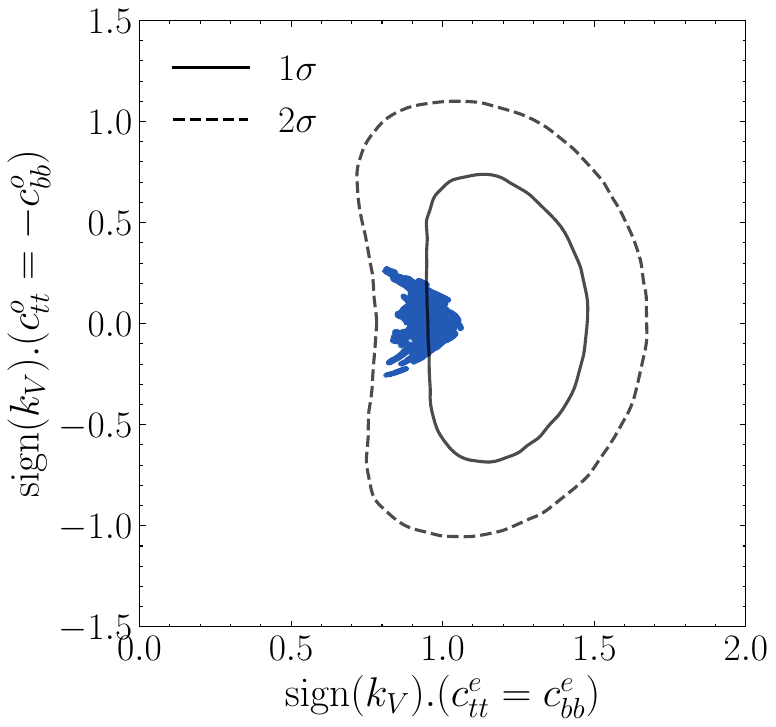}
\vspace*{3mm}
\caption{$h_{125}=h_1$.}
\end{subfigure}
\begin{subfigure}[b]{0.32\linewidth}
\centering
    \includegraphics[width=0.95\textwidth]{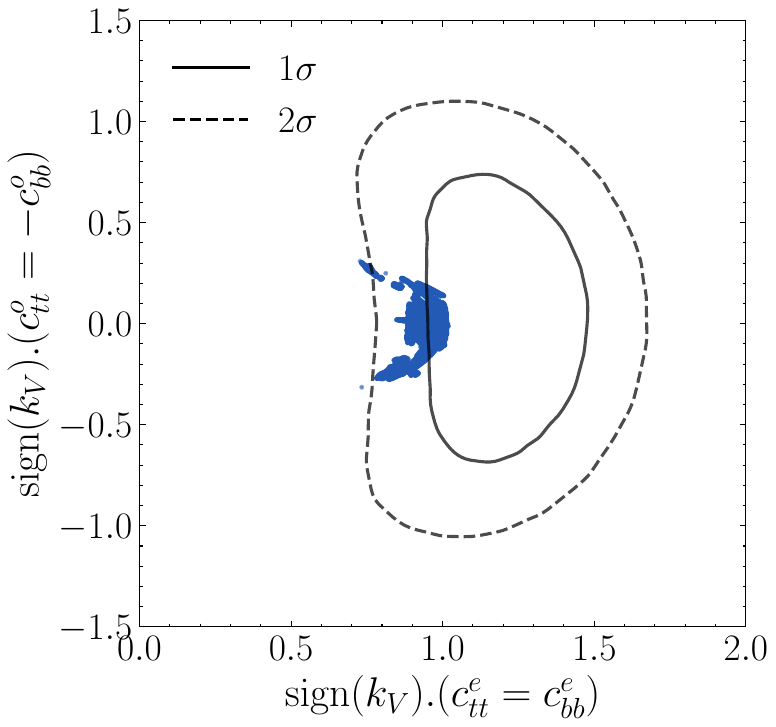}
\vspace*{3mm}
\caption{$h_{125}=h_2$.}
\end{subfigure}
\begin{subfigure}[b]{0.32\linewidth}
\centering
    \includegraphics[width=0.95\textwidth]{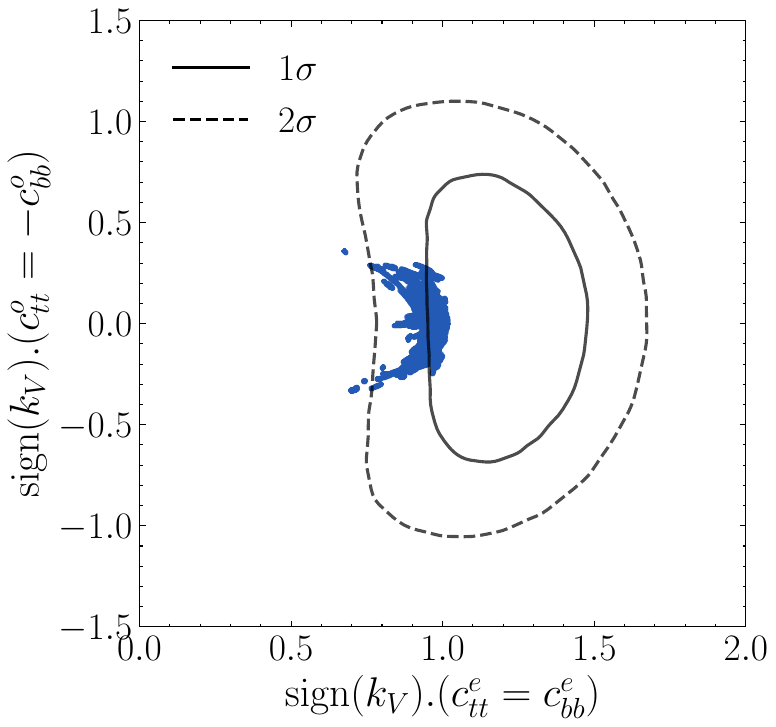}
\vspace*{3mm}
\caption{$h_{125}=h_3$.}
\end{subfigure}
\bigskip
\hspace{2.5cm}
\begin{subfigure}[b]{0.32\linewidth}
\centering
    \includegraphics[width=0.95\textwidth]{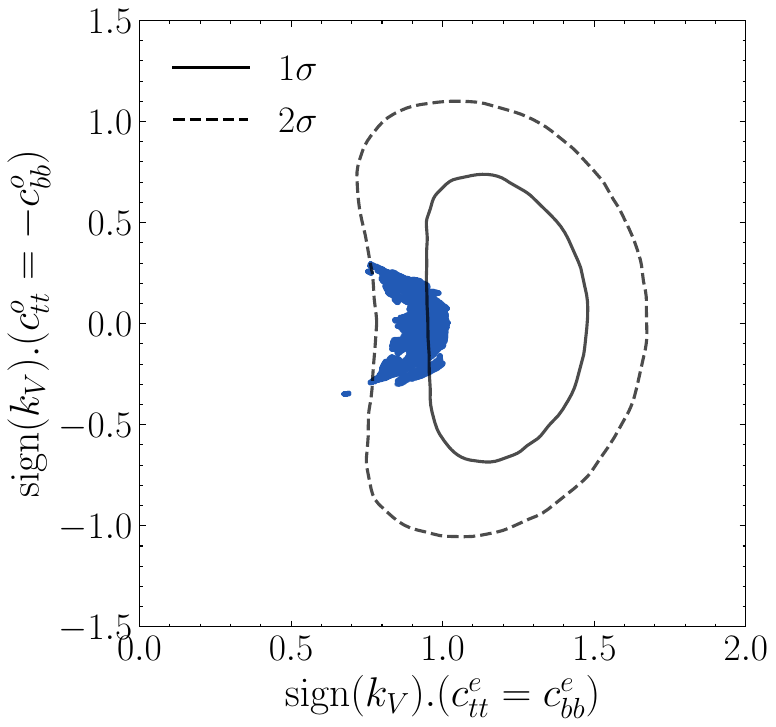}
\vspace*{3mm}
\caption{$h_{125}=h_4$.}
\end{subfigure}
\hfill
\begin{subfigure}[b]{0.32\linewidth}
\centering
    \includegraphics[width=0.95\textwidth]{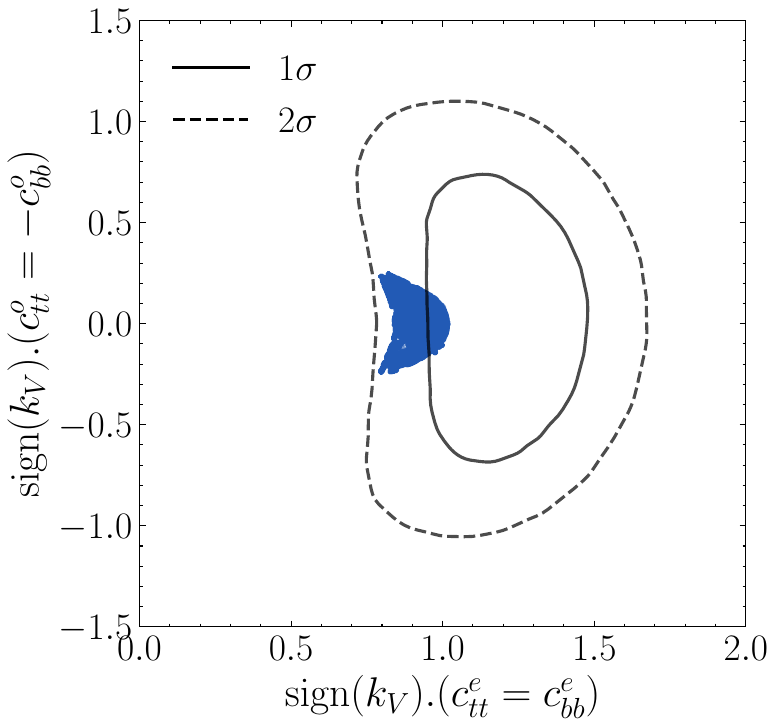}
\vspace*{3mm}
\caption{$h_{125}=h_5$.}
\end{subfigure}
\hspace{2.5cm}
\vspace*{-5mm}
  \caption{Allowed $h_{125}$-top coupling regions in Type X models for the different ordering choices, without novelty reward on the plane shown.
}
  \label{fig:ee_t3_h1,2,3,4,5}
\end{figure}

\begin{figure}[H]\centering
\begin{subfigure}[b]{0.34\linewidth}
\centering
    \includegraphics[width=0.9\textwidth]{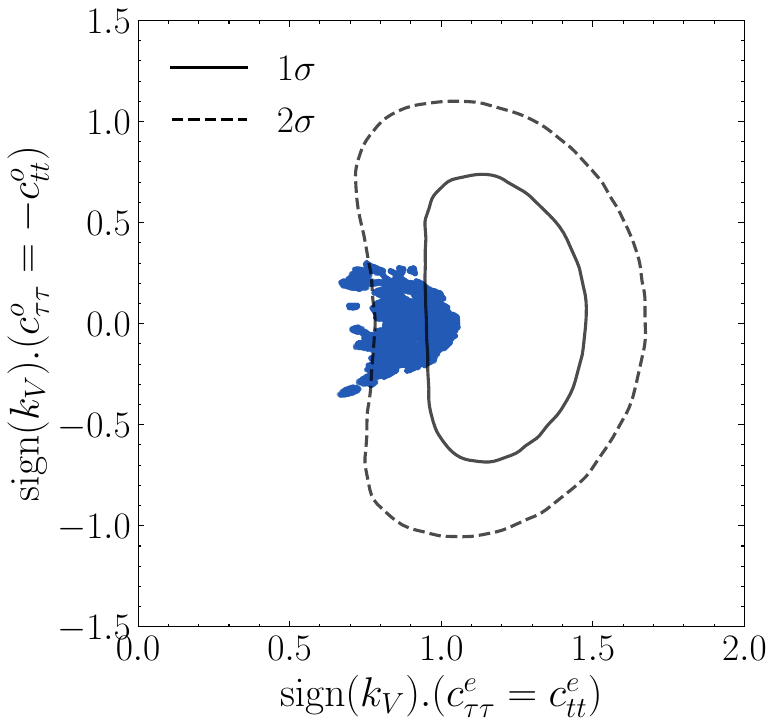}
\vspace*{3mm}
\caption{$h_{125}=h_1$.}
\end{subfigure}
\hspace{2cm}
\begin{subfigure}[b]{0.34\linewidth}
\centering
    \includegraphics[width=0.9\textwidth]{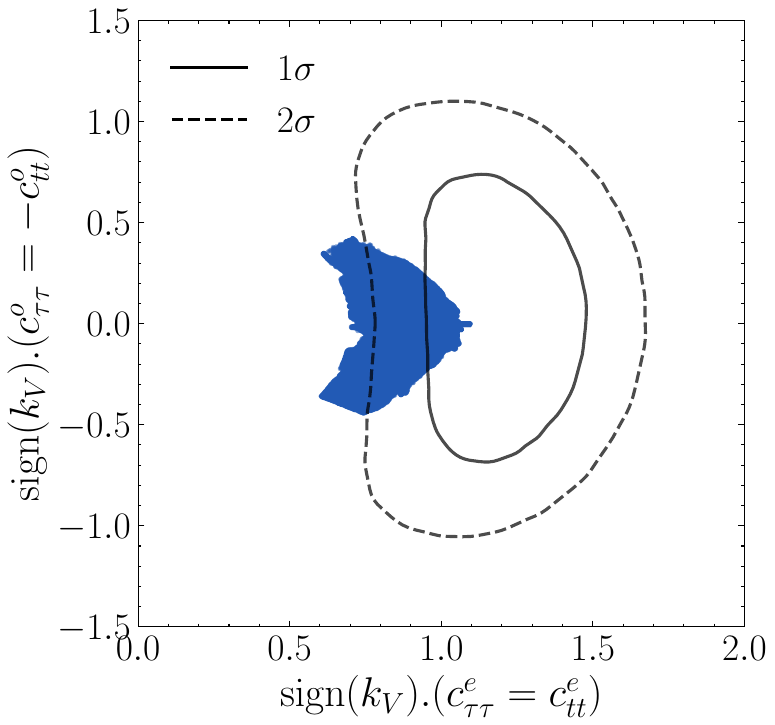}
\vspace*{3mm}
\caption{$h_{125}=h_2$.}
\end{subfigure}
\bigskip
\begin{subfigure}[b]{0.34\linewidth}
\centering
    \includegraphics[width=0.9\textwidth]{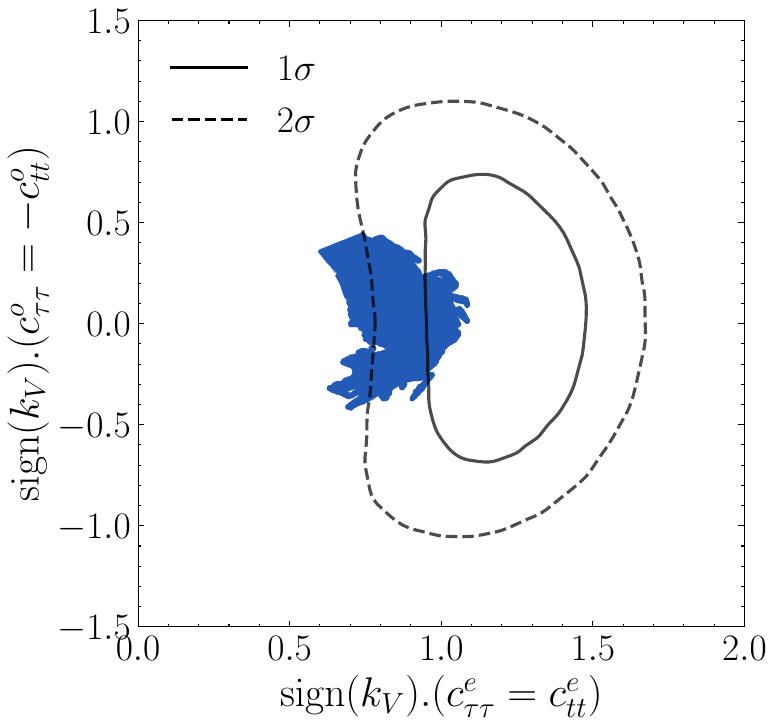}
\vspace*{3mm}
\caption{$h_{125}=h_3$.}
\end{subfigure}
\hspace{2cm}
\begin{subfigure}[b]{0.34\linewidth}
\centering
    \includegraphics[width=0.9\textwidth]{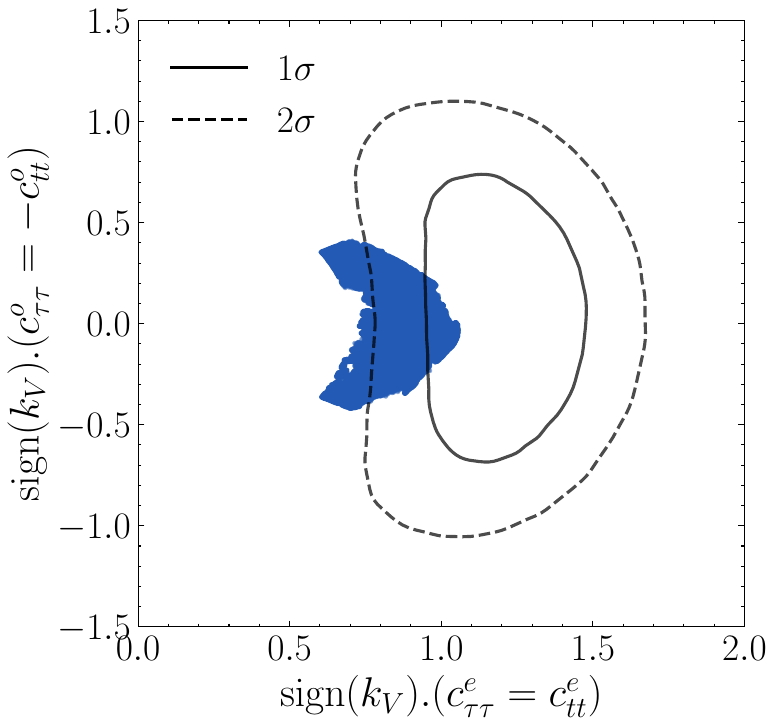}
\vspace*{3mm}
\caption{$h_{125}=h_4$.}
\end{subfigure}

\vspace*{-5mm}
  \caption{Allowed $h_{125}$-top coupling regions in Type Y models, without novelty reward on the plane shown.
}
  \label{fig:tt_t4_h1,2,3,4,5}
\end{figure}

\begin{figure}[H]
  \centering
  \begin{tabular}{ccc}
    \includegraphics[width=0.33\textwidth]{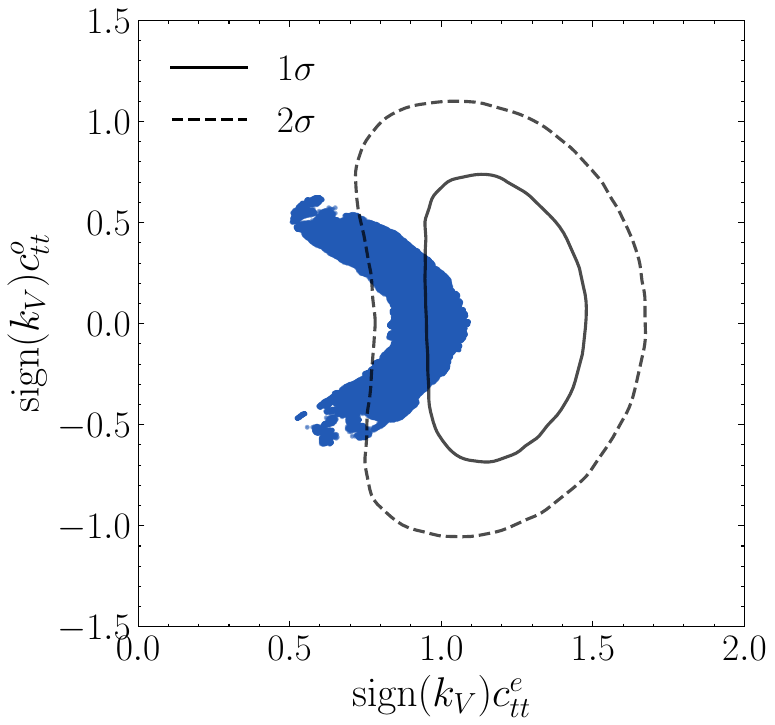}
    &
    \includegraphics[width=0.33\textwidth]{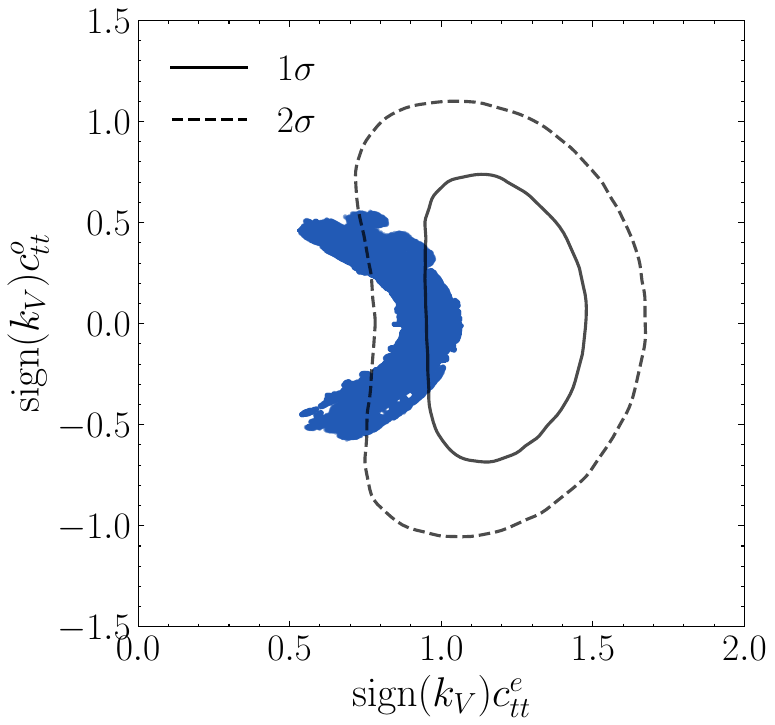}
    &
    \includegraphics[width=0.33\textwidth]{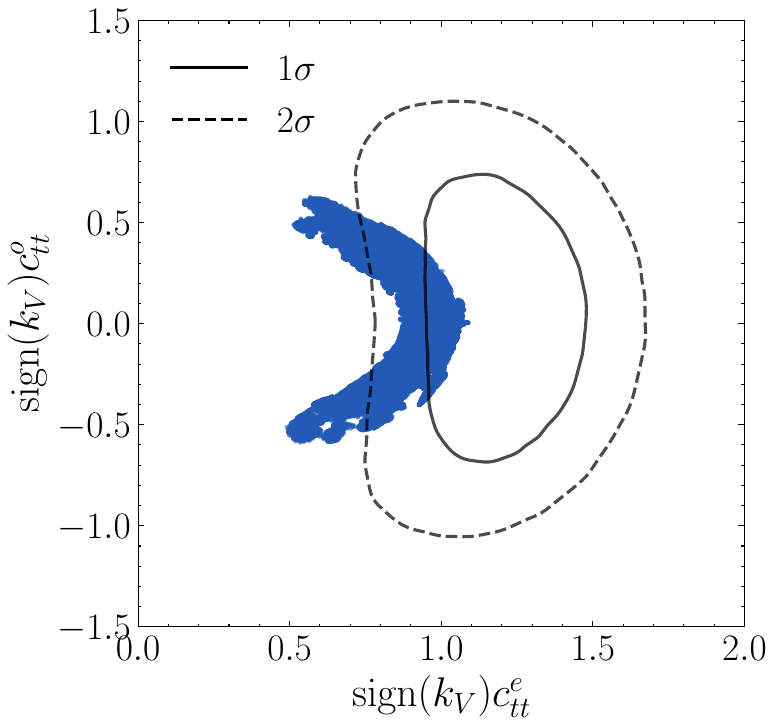}
  \end{tabular}
  \caption{Allowed $h_{125}$-top-type quark coupling regions
in Type Z models with the orderings $h_{125}=h_2$, $h_3$ and $h_4$, from left to right, respectively. The runs shown do not include novelty reward on the plane shown.
%The colour code of the points is the same as in previous plots. 
}
  \label{fig:tt_t5_h2,3,4}
\end{figure}

\section{Benchmarks\label{app:BPs}}

This appendix contains some benchmark points with large pseudoscalar $c^o_{ff}$ terms.
For ease of reference, in this appendix tAhB has the following interpretation:
model A, with A $\in$ (1,2,3,4,5)=(I,II,X,Y,Z);
with the $h_{125}$ corresponding to B=1,2,3,4,5;
respectively.
\begin{table}[H]
\centering
\begin{tabular}{l | c | c | c | c | c | c}
\hline
Parameters & Point 1 t4h2 & Point 2 t4h2 & Point 3 t4h2 & Point 4 t4h2 & Point 5 t2h1 & Point 6 t2h1 \\
\hline
$M_{H_1}$ & 82.9171 & 85.7633 & 71.2235 & 63.0836 & 125.0000 & 125.0000 \\
$M_{H_2}$ & 125.0000 & 125.0000 & 125.0000 & 125.0000 & 316.3615 & 541.5863 \\
$m_{C_1}$ & 264.9851 & 262.4250 & 290.6144 & 247.0139 & 357.2638 & 168.1340 \\
$m_{C_2}$ & 159.8735 & 162.9834 & 144.0619 & 139.8453 & 178.5069 & 437.7242 \\
$\textrm{Re}(m_{12}^2)$ & 4.2426 & 0.2806 & -0.4611 & -3648.3773 & -800.4453 & 0.3778 \\
$\textrm{Re}(m_{13}^2)$ & 0.2366 & 0.6671 & 6974.0226 & 11921.6720 & -17705.1890 & 1157.4426 \\
$\textrm{Re}(m_{23}^2)$ & -7216.2855 & -5925.5554 & -32.0851 & -0.7280 & -38919.0860 & -15123.4290 \\
$\beta_1$ & 0.3098 & 0.2931 & 0.4515 & 0.3274 & 0.4011 & 0.7568 \\
$\beta_2$ & 0.8667 & 0.9439 & 0.7408 & 0.7229 & 1.1256 & 0.7243 \\
$\theta$ & -0.8560 & -0.7944 & -2.1123 & 1.0986 & 2.4258 & 2.5780 \\
$\phi$ & 3.0937 & 3.1183 & -1.4871 & -1.5852 & -3.0573 & 1.4947 \\
$\alpha_{12}$ & -0.7454 & -0.8092 & -2.9066 & -0.0556 & -0.2582 & 0.5948 \\
$\alpha_{13}$ & 0.5201 & 0.4671 & 2.5704 & 0.5611 & 1.0253 & 0.6935 \\
$\alpha_{14}$ & 1.6669 & 1.6460 & -0.2318 & 1.3982 & 0.0823 & -0.0151 \\
$\alpha_{15}$ & 0.5511 & 0.5204 & 1.6743 & 1.8724 & 0.0376 & 0.0744 \\
$\alpha_{23}$ & -2.4994 & -2.3661 & 0.1133 & -0.4400 & 1.6314 & -0.2463 \\
$\alpha_{24}$ & -2.1530 & -2.2429 & 1.5283 & -1.6300 & -2.8859 & 0.6048 \\
$\alpha_{25}$ & -0.0912 & -0.1043 & -1.0026 & -0.3485 & 1.3439 & -1.5145 \\
$\alpha_{34}$ & 0.9160 & 1.0488 & 2.3330 & -0.1000 & -0.0272 & -0.5514 \\
$\alpha_{35}$ & 1.2067 & 1.0604 & -1.7228 & 1.9079 & -1.4227 & 0.2404 \\
$\alpha_{45}$ & -0.0093 & -0.1039 & 1.5361 & -1.6401 & 0.3145 & 0.5447 \\
\hline
$M_{H_3}$ & 153.0237 & 149.3809 & 167.5187 & 148.7508 & 299.8183 & 309.1198 \\
$M_{H_4}$ & 250.0512 & 244.3027 & 244.0183 & 196.3440 & 246.6197 & 283.9242 \\
$M_{H_5}$ & 150.5510 & 147.3391 & 177.6647 & 153.8856 & 481.9641 & 323.8520 \\
\hline
$c^e_{tt}$ & 0.9102 & 0.9073 & 0.7494 & 0.8189 & 0.9433 & 0.9619 \\
$c^o_{tt}$ & 0.0669 & 0.0733 & -0.3646 & 0.1890 & -0.0292 & -0.0630 \\
$c^e_{bb}$ & -0.2627 & -0.3104 & 0.5040 & -0.3967 & -0.7848 & 0.8355 \\
$c^o_{bb}$ & 0.7034 & 0.7054 & -0.5162 & 0.5009 & 0.2360 & -0.1220 \\
$c^e_{\tau\tau}$ & 0.9102 & 0.9073 & 0.7494 & 0.8189 & -0.7848 & 0.8355 \\
$c^o_{\tau\tau}$ & -0.0669 & -0.0733 & 0.3646 & -0.1890 & 0.2360 & -0.1220 \\
\hline
%\hline
\end{tabular}
\caption{Parameter values for benchmark points from the scans.
The table shows masses, mixing angles, and Higgs couplings to fermions.
Points 1-4 are taken from the t4h2 scan; see Fig.~\ref{fig:ttvsbb_t4_h2}.
Points 5-6 are taken from the t2h1 scan; see Fig.~\ref{fig:ee_t2_h1,2,3,4}(a).
\label{table:benchmark_point_comprehensive}}
\end{table}
The points selected have passed all current constraints.
They have the following characteristics:
\begin{itemize}
\item Points 1 and 2 (Type Y) have maximal $c^0_{bb}$ despite having a very small
$|c^0_{tt}|=|c^0_{\tau\tau}|$.
They correspond to the very exciting possibility that, even faced with all current constraints,
$h_{125}$ can couple to top and tau as almost pure scalar, while coupling to the bottom
as almost pure pseudoscalar.
\item Points 3 and 4 (Type Y) have a large $c^0_{bb}$ and a $|c^0_{tt}|=|c^0_{\tau\tau}|$
large enough that these points can be excluded in the near future with a modest
increase in the precision in Eqs.~\eqref{theta_t}-\eqref{theta_tau}.
\item Points 5 and 6 (Type II) have $|c^0_{bb}|=|c^0_{\tau\tau}|$, corresponding to regions
between the $1\sigma$ and $2\sigma$ lines of Eq.~\eqref{theta_tau}.
They are thus, currently allowed, but can also be falsified soon.
\end{itemize}
The conclusion is that modest increases in the experimental determination
of $\theta_t$ and/or $\theta_\tau$ can have a dramatic impact on this model.
On the other hand, it is also possible that an almost pure
CP-odd $hbb$ coupling (with almost pure CP-even $htt$ and $h\tau\tau$ couplings)
can survive further scrutiny.

%%%%%%%%%%%%%%%%%%%%%%%%%%%%%%%%%%%%%%%%%%%%%%%%%%%%%%%%%%%%%%%%%%%
%%%%%%%%%%%%%%%%%   References %%%%%%%%%%%%%%%%%%%%%%%%%%%%%%%%%%%%
%%%%%%%%%%%%%%%%%%%%%%%%%%%%%%%%%%%%%%%%%%%%%%%%%%%%%%%%%%%%%%%%%%%
\bibliographystyle{JHEP}
%\bibliography{BFB.bib}
\bibliography{Bibliography}

\end{document}